\documentclass[a4paper,fleqn,usenatbib,useAMS]{mnras}
\usepackage{aas_macros}
\usepackage{graphicx}
\usepackage{multirow,psfig}
\usepackage{graphicx}	
\usepackage{amsmath}	
\usepackage{amssymb}	
\usepackage{color,natbib}
\usepackage{pgf}

\title[PG-RQS. I. The 45-GHz observations.] {The PG-RQS survey. Building the
  radio spectral distribution of radio-quiet
  quasars.~I. The 45-GHz data.} \author[R.~D. Baldi et al.]
      {R. D. Baldi$^{1,2}$\thanks{E-mail: ranieri.baldi@inaf.it},
        A. Laor$^{3}$, E. Behar$^{3}$, A. Horesh$^{4}$,
        F. Panessa$^{5}$, I. McHardy$^{2}$, \newauthor
        A. Kimball$^{6}$\\ $^{1}$ INAF - Istituto di Radioastronomia, Via
        P. Gobetti 101, I-40129 Bologna, Italy; \\ $^{2}$ School of
        Physics and Astronomy, University of Southampton, Southampton,
        SO17 1BJ, UK;\\ $^{3}$ Department of Physics, Technion 32000,
        Haifa 32000, Israel\\ $^{4}$ Racah Institute of Physics,
        Hebrew University of Jerusalem, Jerusalem 91904,
        Israel\\ $^{5}$ INAF - Istituto di Astrofisica e Planetologia
        Spaziali, via Fosso del Cavaliere 100, I-00133 Roma,
        Italy.\\ $^{6}$ National Radio Astronomy Observatory, 1003
        Lopezville Rd, Socorro, NM 87801, USA}
\begin{document}


\pagerange{\pageref{firstpage}--\pageref{lastpage}} \pubyear{2018}

\maketitle

\label{firstpage}

\begin{abstract}


The origin of the radio emission in radio-quiet quasars (RQQs) remains
unclear. Radio emission may be produced by a scaled-down version of the
relativistic jets observed in radio-loud (RL) AGN, an AGN-driven wind,
the accretion disc corona, AGN photon-ionisation of ambient gas
(free-free emission), or star formation (SF). Here, we report a pilot
study, part of a radio survey (`PG-RQS') aiming at exploring the
spectral distributions of the 71 Palomar-Green (PG) RQQs: high angular
resolution observations ($\sim$50 mas) at 45~GHz (7 mm) with the Karl G.
Jansky Very Large Array of 15 sources. Sub-mJy radio cores are
detected in 13 sources on a typical scale of $\sim$100 pc, which
excludes significant contribution from galaxy-scale SF. For 9 sources
the 45-GHz luminosity is above the lower
frequency ($\sim$1--10 GHz) spectral extrapolation, indicating the
emergence of an additional flatter-spectrum compact component at high
frequencies. The X-ray luminosity and black hole (BH) mass, correlate
more tightly with the 45-GHz luminosity than the 5-GHz.  The
45\,GHz-based radio-loudness increases with decreasing Eddington ratio
and increasing BH mass M$_{\rm BH}$.  These results suggest that the 45-GHz
emission from PG RQQs nuclei originates from the innermost region of
the core, probably from the accretion disc corona. Increasing
contributions to 45-GHz emission from a jet at higher M$_{\rm BH}$ and
lower Eddington ratios and from a disc wind at large Eddington ratios
are still consistent with our results. Future full radio spectral
coverage of the sample will help us investigating the different
physical mechanisms in place in RQQ cores.

\end{abstract}


\section{Introduction}
\label{intro}

The origin of the radio emission in Radio-loud Active Galactic Nuclei
(RL AGN) is clear, luminous relativistic jets of magnetised plasma,
which can extend far out, on the host galaxy scale and
beyond. Conversely, radio-quiet (RQ) AGN are associated with radio
emission which is typically $10^3$ times weaker \citep[as defined
  by][]{kellermann89}, in smaller structures (kpc-pc scale,
e.g. \citealt{blundell96,nagar99,ulvestad05_n4151,gallimore06}) with
sub-relativistic velocities
(e.g. \citealt{middelberg04,ulvestad05_rqq}) compared to RL AGN. The
reduced sizes and low brightness of the radio emission of RQ AGN
create a major challenge for detailed studies, in sharp contrast with
the thoroughly studied RL AGN. The fewer radio studies of RQ AGN
(e.g. \citealt{barvainis89,kellermann94,barvainis96,kukula98,ulvestad05_n4151,leipski06,padovani11,doi11,zakamska16,jarvis19,smith20b,fawcett20,nyland20,jarvis21,baldi21a})
generally lead to mixed results.  This encourages to keep
investigating it, as it indicates that the origin of the radio
emission in RQ AGN is still an open question. If a number of different
processes are indeed involved, then the radio band can be used to
probe a range of physical processes, rather than being heavily
dominated by a single process, as occurs in RL AGN (see
\citealt{panessa19,blandford19} for reviews). From large scale to
small scale: i) {\it host galaxy star formation} could account for the
observed FIR-to-radio emission observed in active and non active
galaxies \citep{condon13,zakamska16}; ii) an {\it AGN-driven wind} is
expected to shock the interstellar gas, leading to particle
acceleration and radio synchrotron emission, which may reach the
observed flux level \citep{jiang10}; iii) the intense radiation of the
AGN photo-ionizes large volumes of ambient gas, as supported by the
strength of the narrow and broad line emission observed in type-1 AGN,
leading to {\it thermal free-free emission} in the radio band
\citep{baskin21}; iv) a {\it scaled-down jet}, physically similar to
the one in RL AGN, but much fainter, less energetic and slower
\citep{barvainis96, gallimore06,talbot21}; v) a tight radio/X-ray
luminosity relation for AGN ($\sim$10$^{-5}$, \citealt{laor08})
similar to coronally active stars \citep{guedel93,guedel02} suggests
that {\it coronal emission} from magnetic activity above the accretion
disc \citep{field93,gallimore97} may produce the observed radio
emission.

High resolution and high sensitivity radio observations are
fundamental to determine the physical mechanism which produces the
radio emission in RQ AGN.  The detection of a low-brightness radio
core at the center of the radio structures, where the supermassive
black hole (BH) resides, requires observations which can resolve the
optically-thick synchrotron source, whose size scales as $\nu^{-5/4}$
$L^{1/2}$ (e.g. \citealt{laor08,inoue18}). Therefore, radio
observations at higher-frequencies, which probe the emission on
smaller scales, are required. Nevertheless, since the typical radio
spectra of RQ AGN are steep in the cm-band, the detection rate could
benefit from observations at intermediate radio frequencies, between
the cm and mm bands. In particular, in the band, 2--0.6 cm
(15-50~GHz), the probed physical scale of the radio-emitting region in
nearby AGN is $\sim$0.01 pc, which is an order of magnitude smaller
than the optically-thick 5-GHz emission region.

As opposed to observations at 1.4-8.5 GHz, higher frequency
observations of RQ AGN in the range 15-50 GHz are scarce. In the last
decade, more efforts have been spent towards studying the radio
emission from RQ AGN moving gradually from cm band to shorter
wavelengths (e.g.,
\citealt{doi05,murphy10,park13,inoue14,behar15,smith16,behar18,ricci19,smith20}).

\begin{figure*}
\includegraphics[angle=90,width=0.9\textwidth,height=0.37\textwidth]{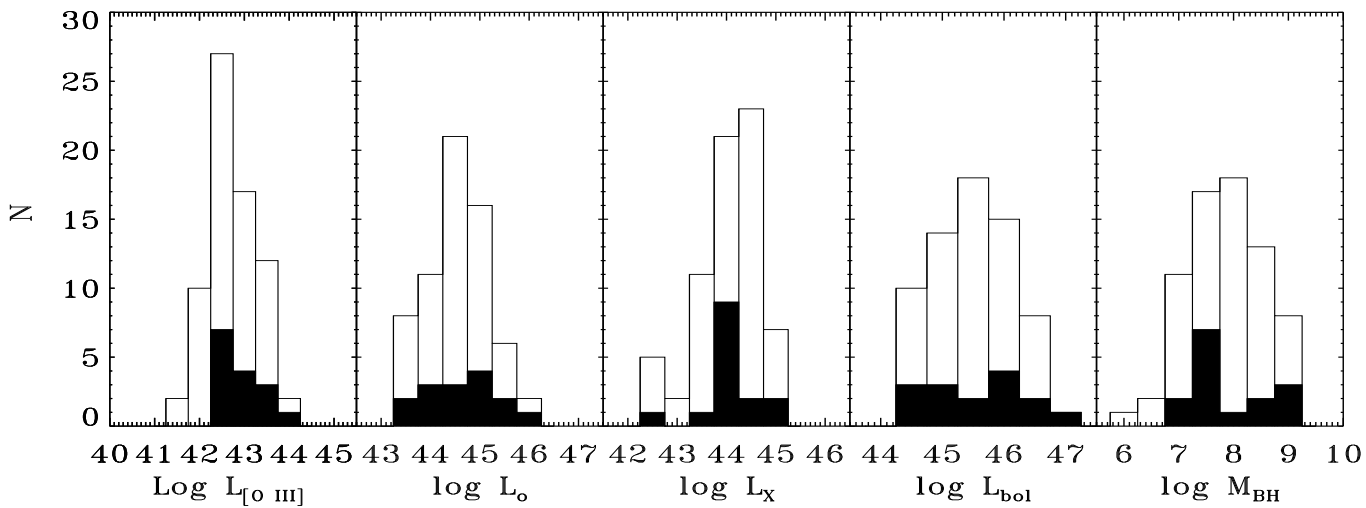}
\label{histo}
\caption{Distribution of the [O~III] line, optical (4400 \AA), X-ray (2-10 keV), bolometric
  luminosities (in erg s$^{-1}$) and BH masses (in M$_{\odot}$) of the whole PG RQQ sample (71
  objects). The filled histograms represent the distribution of the 15
  PG RQQs studied in this work. The KS tests confirm that the 15 sources are
  representative of the entire PG RQQ sample.}
\end{figure*}

Radio observations at the cm-band with sub-mJy sensitivity and
sub-arcsecond resolution have found extended emission on kpc scale or
compact unresolved emission in the majority of RQ AGN (e.g.,
\citealt{nagar02,giroletti09,panessa13,maini16,baldi18,chiaraluce19,baldi21a}).
The flat or even inverted spectrum observed at higher radio
frequencies indicates a compact optically-thick source
\citep{doi05,doi11,behar15,behar18}, superimposed on the
steep-spectrum which originates from more extended optically-thin
structures that dominate at lower frequencies. Such a flat-spectrum
core is usually identified as the base of a possible jet or outflow,
and may be located in the X-ray corona \citep{markoff05}.  To
  explore the nature of the radio core, the identification of such a component, resolved
  with high resolution observations, may help us to carry on an
  unbiased comparison with the well-studied cores of RL AGN. On one
  hand, empirical relations between the radio core luminosity and
  global properties of the AGN/galaxy have been found for RL AGN
  (e.g. radio vs BH or galaxy mass, colour, SF, Eddington ratio,
  see \citealt{heckman14}). On the other hand, for RQ AGN a similar
  study is limited because a few studies on arcsec- and mas-scale
  radio emission have led to conflicting or inconclusive results (e.g. the presence or
  not of a radio-M$_{\rm BH}$ relation,
  e.g. \citealt{lacy01,mclure04,bian08,panessa13,baldi21b}). Therefore,
  dedicated high-resolution radio observations on large samples of RQ
  AGN are needed to resolve their cores, and enable us to perform an impartial
  comparison with RL AGN and their radio-based empirical relations.

Quasars, being luminous AGN, are perfect settings to explore the
different mechanisms of radio emission.  Searching for a connection
between accretion/ejection properties and the radio emission in RQQs
is, thus, the goal of our PG-RQS (Palomar-Green Radio-Quiet SED)
survey by building up the radio (from MHz to mm band) spectral energy
distributions of the 71 Palomar-Green (PG) RQQs from the 87 Palomar
Bright Quasar Survey at $z<0.5$ \citep{schmidt83,boroson92}. The PG
survey was the first large solid angle (10,000 square degrees)
complete quasar survey, selected to be blue and point-like, and it
remains the largest complete survey of optically bright quasars at $B
<$ 16, and therefore has no radio selection biases. It therefore
provides an unbiased view of the distribution of radio emission
properties of RQ AGN. The PG RQQ sample represents the most
extensively studied sample of type-1 AGN and, thus, the cornerstone
for quasar studies in the last decades: including, e.g. the overall
SED \citep{neugebauer87,sanders89}, radio continuum and imaging
\citep{kellermann89,kellermann94,miller93}, infrared photometry
\citep{haas03,shi14,petric15}, optical spectroscopy \citep{boroson92},
optical polarisation \citep{berriman90}, UV spectroscopy
\citep{baskin05}, soft X-ray spectroscopy \citep{brandt00}, and radio
spectral indices \citep{laor19}. From these studies, various
fundamental properties were derived, such as M$_{\rm BH}$ (from broad
H$\beta$ emission line widths, \citealt{kaspi00,vestergaard06}, and from
stellar velocity dispersions, \citealt{tremaine02}), accretion rates
and radiative efficiencies \citep{davis11}, and numerous other
observed properties.  This sample revealed various interesting trends,
such as the eigenvector 1 (EV1) set of emission line correlations
\citep{boroson92}, the M$_{\rm BH}$ -- bulge mass relation in AGN
\citep{laor98}, and the M$_{\rm BH}$ -- radio loudness relation
\citep{laor00}. In conclusion, for its optical selection and its
  large multi-band coverage, the PG RQQs represents an ideal sample to
  search for radio-based empirical relations in the RQ
  regime. Although characterised by relatively high accretion rates
  (in Eddington units), the extreme conditions of the accretion and
  ejection of the PG RQQs could help in the identification of the
  origin of the radio emission among the different physical
  mechanisms.  Precisely, in this work we present the results from a
pilot study of 15 PG RQQs observed at 45 GHz with the Karl G. Jansky
Very Large Array (VLA).

This paper is organised as follows. In Sect.~\ref{sec:sample} we
present the project, the PG sample, and the
observations. Sect.~\ref{results} shows the main radio properties of
the sample and we seek for multi-band correlations in
Sect~\ref{multicorr}. A comparison of the results with RLQs is in
Sect.~\ref{comparison} to highlight the similarities and differences
with the radio properties of RQQs, which we discuss in
Sect~\ref{sec:discussion}. We elaborate the final interpretation of
the results based on the different radio production processes in RQQs
in Sect.~\ref{origin} and on the comparison with X-ray Binary systems
in Sect.~\ref{XRB}. We draw our conclusions in Sect~\ref{concl}.

\begin{table*}
\begin{center}
  \caption[Properties of the RQQ sample.]{VLA A-array 45-GHz 
    properties of the observed 15 PG RQQs.}
\begin{tabular}{ll|cccccccc}
\hline 
PG name & alternative   & z &  size & F$_{\rm core}$    &   F$_{\rm tot}$  & rms                 & uvtaper   &  $F_{uvtaper}$  &  P$_{45~{\rm GHz}}$ \\
       &  name          &   &  pc  & mJy beam$^{-1}$  & mJy             &    mJy beam$^{-1}$  & k$\lambda$ & mJy beam$^{-1}$ & erg s$^{-1}$\\
 (1)   &  (2)           &(3)& (4)  & (5)              & (6)             &    (7)              & (8)        & (9)             & (10)        \\
\hline
PG~0003+199 & MRK~335  & 0.0258 &  52 & 0.50$\pm$0.04   &  0.60   &  0.036 &  300   & 0.82$\pm$0.08  & 38.52  \\
PG~0026+129 &          & 0.1420 & 251 & 0.20$\pm$0.04   &  0.21   &  0.037 &  300   & 0.53$\pm$0.11  & 39.71   \\
PG~0049+171 & MRK1148  & 0.0640 & 124 & 0.63$\pm$0.05   &  0.64   &  0.040 &  300   & 0.81$\pm$0.10  & 39.46   \\
PG~0050+124 & UGC~545  & 0.0589 & 115 & $<$0.15         &         &  0.051 &  500   & 0.30$\pm$0.08  & 39.06 \\
PG~0052+251 &          & 0.1544 & 270 & 0.37$\pm$0.04   &  0.37   &  0.035 &  300   & 0.62$\pm$0.08  & 40.05   \\
PG~0157+001 & MRK~1014 & 0.1631 & 282 & 0.48$\pm$0.06   &  0.52   &  0.058 &  300   & 0.55$\pm$0.11  & 40.22    \\                 
PG~0844+349 &          & 0.0640 & 124 & $<$0.06         &         &  0.020 &  300   &  $<$0.19       & $<$38.46 \\
PG~1001+054 &          & 0.1611 & 279 & $<$0.06         &         &  0.019 &  250   &  0.13$\pm$0.04 & 39.65\\
PG~1048+342 &          & 0.1670 & 288 & 0.17$\pm$0.03   &  0.24   &  0.033 &  300   &  0.41$\pm$0.12 & 39.80   \\
PG~1049-005 &          & 0.3599 & 508 & $<$0.11         &         &  0.035 &  600   & 0.25$\pm$0.07  & 40.72\\
PG~1114+445 &          & 0.1439 & 254 & 0.13$\pm$0.02   &  0.16   &  0.025 &  600   & 0.17$\pm$0.05  & 39.54   \\
PG~1149-110 &          & 0.0490 &  96 & 0.30$\pm$0.03   &  0.36   &  0.032 &  600   & 0.31$\pm$0.06  & 38.93   \\   
PG~1229+204 & MRK~771  & 0.0630 & 122 & 0.27$\pm$0.03   &  0.27   &  0.028 &  300   & 0.36$\pm$0.08  & 39.12   \\
PG~1259+593 & SBS~1259+593 & 0.4778 & 602 & $<$0.09      &         &  0.029 &  300   & $<$0.24        & $<$40.57 \\
PG~1310-108 & II~SZ~010 & 0.0343 &  69 & 0.19$\pm$0.03   &  0.24   &  0.039 &  300   & 0.38$\pm$0.07  & 38.42  \\
\hline                                          
\end{tabular}                                   
\label{lum}
\begin{flushleft}
  Column description: (1)-(2) PG source name and alternative name; (3)
  redshift; (4) physical size of 100~mas; (5)--(6) 45-GHz peak flux densities (mJy beam$^{-1}$, derived from CASA \texttt{`imfit'} task)
  and total flux densities (mJy) from full-resolution maps; (7) rms
  (mJy beam$^{-1}$); (8) $uv$taper scale parameter; (9) 45-GHz  peak flux density
  measured in low-resolution map (mJy beam$^{-1}$); (10) 45-GHz core
  luminosity (erg s$^{-1}$) from  F$_{\rm core}$ . Upper limits are evaluated at 3$\sigma$
  level.
\end{flushleft}
\end{center}
\end{table*}

\begin{figure*}
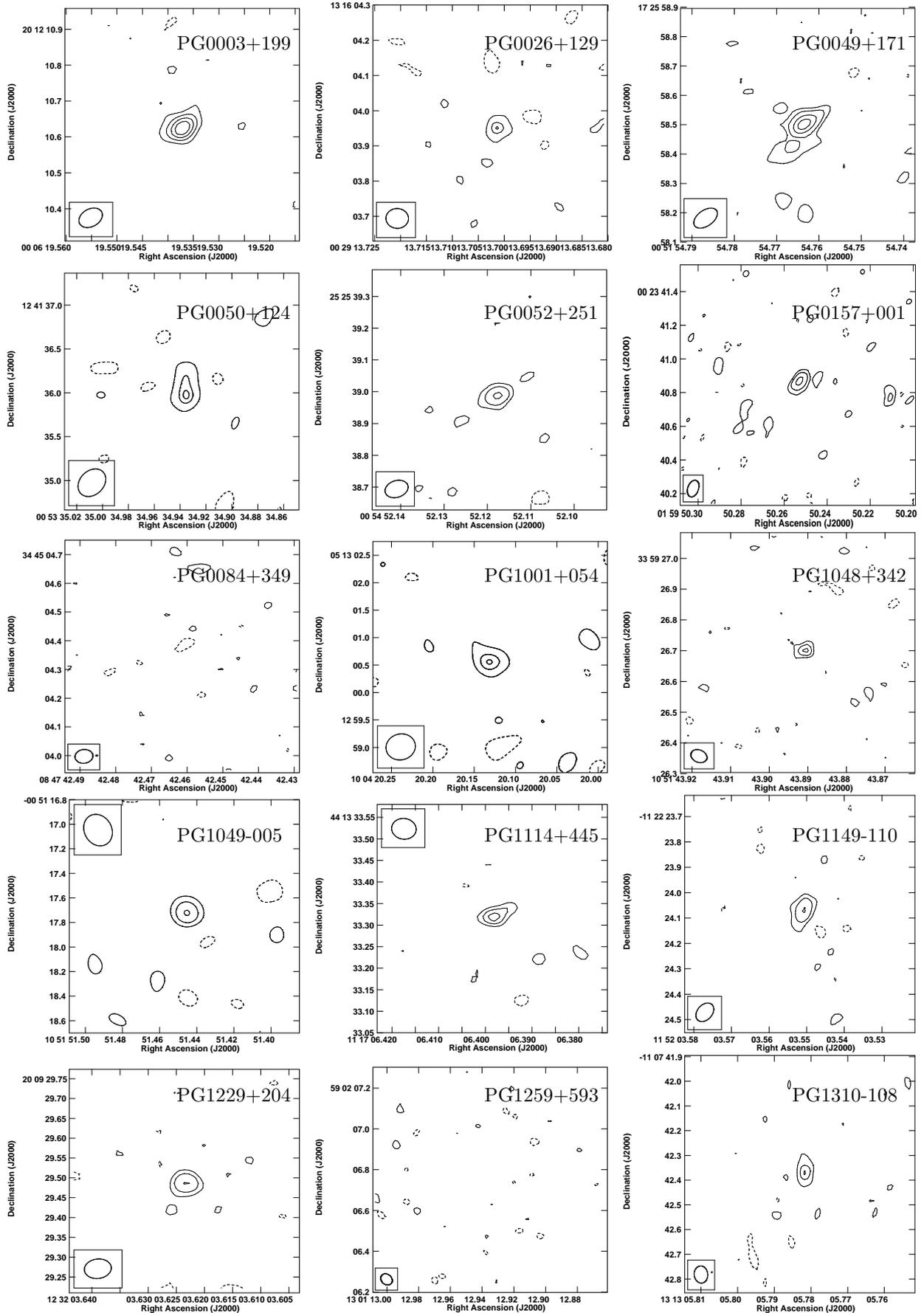

  \vspace{-0.3cm}
  \centerline{
    \put(90,110){{\large PG0003+199}}
 \includegraphics[width=0.65\columnwidth]{mrk335.ps}
  \put(90,110){{\large PG0026+129}}
  \includegraphics[width=0.65\columnwidth]{PG0026.PS}
  \put(90,110){{\large PG0049+171}}
  \includegraphics[width=0.65\columnwidth]{mrk1148.ps}}
  \vspace{-0.2cm}
\centerline{
    \put(90,110){{\large PG0050+124}}
 \includegraphics[width=0.65\columnwidth]{ugc545uvtap.ps}
  \put(90,110){{\large PG0052+251}}
  \includegraphics[width=0.65\columnwidth]{pg0052.ps}
  \put(90,110){{\large PG0157+001}}
  \includegraphics[width=5.5cm,height=5cm]{mrk1014.ps}}
  \vspace{-0.2cm}

\centerline{
    \put(90,110){{\large PG0084+349}}
 \includegraphics[width=0.65\columnwidth]{pg0844.ps}
  \put(90,110){{\large PG1001+054}}
  \includegraphics[width=0.65\columnwidth]{pg1001uvtap.ps}
  \put(90,110){{\large PG1048+342}}
  \includegraphics[width=0.65\columnwidth]{pg1048.ps}}
  \vspace{-0.2cm}

\centerline{
    \put(90,110){{\large PG1049-005}}
 \includegraphics[width=0.65\columnwidth]{pg1049uvtap.ps}
  \put(90,110){{\large PG1114+445}}
  \includegraphics[width=0.65\columnwidth]{PG1114.ps}
  \put(90,110){{\large PG1149-110}}
  \includegraphics[width=0.65\columnwidth]{PG1149.ps}}
  \vspace{-0.2cm}

\centerline{
    \put(90,110){{\large PG1229+204}}
 \includegraphics[width=0.65\columnwidth]{mrk771.ps}
  \put(90,110){{\large PG1259+593}}
  \includegraphics[width=0.65\columnwidth]{SBS1259.ps}
  \put(90,110){{\large PG1310-108}}
\includegraphics[width=0.65\columnwidth]{iisz010.ps}}
  \vspace{-0.3cm}
\caption{The A-array VLA 45-GHz maps of the 15 PG RQQs in full
  resolution (40-80 mas), except for PG~0050+124,PG~1001+054,
  PG~1049-005 for which we present the $uv$-tapered map where the core
  emission is detected at a lower resolution ($<$0.5 arcsec). 
Two objects, PG~0844+349 and PG~1259+593, are not detected.
The map properties (beam
and contour levels) are presented in Table~\ref{tab_cont}.}
 \label{maps}
\end{figure*}

\begin{table}
\begin{center}
  \caption[Properties of the RQQ sample.]{Parameters of the JVLA 45-GHz
    maps (Fig.~\ref{maps}).}
\begin{tabular}{l|ccc}
\hline 
PG name & beam   & PA  & contour levels  \\ 
\hline
PG~0003+199 &   0.069$\times$0.049 & -60.6 &  0.09.5$\times$(-1,1,2,3,4)   \\ 
PG~0026+129 &    0.064$\times$0.056 &  87.3 &  0.080$\times$(-1,1,2,2.4)   \\
PG~0049+171 &   0.090$\times$0.053 &  -54.3 &  0.085$\times$(-1,1,2,4,6)  \\  
PG~0050+124 &   0.35$\times$0.27   &  -44.5 & 0.25$\times$(-1,1,2,2.5)  \\   
PG~0052+251 &   0.074$\times$0.052 &  -71.5 &  0.080$\times$(-1,1,2,4) \\    
PG~0157+001 &   0.10$\times$0.073  & -28.5 &  0.10$\times$(-1,1,2,4) \\        
PG~0844+349 &  0.062$\times$0.048 &  -83.4 & 0.047$\times$(-1,1)  \\        
PG~1001+054 &   0.53$\times$0.47 & -75.6 &  0.050$\times$(-1,1,2,2.5) \\
PG~1048+342 &   0.054$\times$0.040 &   67.2 &  0.077$\times$(-1,1,1.5,2) \\ 
PG~1049-005 &   0.27$\times$0.22   &   28.2 &  0.12$\times$(-1,1,1.5,2)  \\    
PG~1114+445 &   0.057$\times$0.048 &  82.2  & 0.055$\times$(-1,1,1.5,2) \\   
PG~1149-110 &  0.083$\times$0.058 &  -41.2 & 0.075$\times$(-1,1,2,4)  \\    
PG~1229+204 &   0.068$\times$0.049  &  -82.7 & 0.065$\times$(-1,1,2,4)  \\   
PG~1259+593 &  0.061$\times$0.050 &  48.5 &  0.070$\times$(-1,1) \\           
PG~1310-108 &   0.070$\times$0.054 &    5.3  & 0.065$\times$(-1,1,2,2.8) \\    
\hline                                          
\end{tabular}                                   
\label{tab_cont}
\begin{flushleft}
Column description: (1) source name; (2) FWHM of the elliptical
Gaussian restoring beam (in arcsec) of the maps presented in
Fig.~\ref{maps}; (3) PA of the restoring beam (degree); (4) radio
contour levels (mJy beam$^{-1}$).
\end{flushleft}
\end{center}
\end{table}

\section{Sample and Observations}
\label{sec:sample}

In this work we report the A-array VLA observations for a sub-sample
of 15 objects (see Table~\ref{lum}), randomly selected from the whole
71 RQ PG quasars at z$<$0.5 \citep{boroson92}. The PG sample covers
AGN over a wide range of intrinsic properties, on a wide interval of
bolometric luminosities L$_{\rm Bol}$ $\sim$10$^{43}$-10$^{46}$ erg
s$^{-1}$ and with a M$_{\rm BH}$ ranging from a few 10$^6$ M$_{\odot}$
to a few 10$^9$ M$_{\odot}$. The wealth of additional information
available for this sample allows to explore possible relations between
the 45\,GHz emission and other emission and absorption properties,
which are known to be related (e.g. \citealt{boroson92}).

Figure~\ref{histo} depicts the distributions of the main AGN luminosity
estimates ([O~III] emission line L$_{\rm [O~III]}$, 4400 \AA\, optical  L$_{\rm o}$, 2-10 keV X-ray L$_{\rm X}$, bolometric luminosities\footnote{Bolometric luminosities of the PG QSOs are calculated via broad-band (optical, UV and X-ray) SED modeling by \citet{davis11}.}
  L$_{\rm bol}$ and BH masses M$_{\rm BH}$ taken from
  \citet{boroson92}, \citet{laor08} and \citet{davis11} for all 71
  PG RQQs. The solid histograms show the distributions of such
  parameters for the 15 PG RQQs studied here. We test if our pilot
  sample is consistent with the hypothesis that it is randomly drawn from the
  distribution of the whole sample using the Kolmogorov–Smirnov test
  (KS test). The KS-test probability for each distribution is $>$90
  per cent and confirm that 15 targets are representative of the
  entire PG RQQ sample.

  This pilot observations of 15 PG RQQs at 45 GHz with the VLA is the
  first of a series of studies, part of the PG-RQS survey, aiming at
  exploring the PG RQQs at high and low radio frequencies with
  different radio arrays and angular resolutions: at arcsec resolution
  with {\it GMRT} and {\it LOFAR} in the MHz regime to milli-arcsec
  resolution with very-long baselined interferometry (VLBI)
  observations (such as {\it VLBA} and {\it eMERLIN}), and including
  high frequency observations with VLA (1--45 GHz) and with {\it
    ALMA} (85--500 GHz). An accurate determination of the broad-band
  SEDs of PG RQQs will allow us to probe the radial distribution of
  the radio emission at different scales (set by the band frequency
  and resolution), and constraints possible emission mechanisms.
  
The VLA observations of the pilot sample presented in this work were carried out in Q-band
with 8-GHz bandwidth with exposure times of 12 minutes on source
between October 2016 and January 2017 (project 16B-126). Flux and
phase calibrator were observed, respectively, at the beginning of each
scan (4 min) and in cycles with the target (6 min).

We used the CASA pipeline to calibrate the visibility data, including
the flagging process.  After obtaining the phase and amplitude
solution using the calibrators, we applied them to the target source
using standard procedures. Eventually, we inverted and cleaned with
natural weight the visibility data to obtain the final images with an
angular resolution of $\sim$40-80 mas and a typical rms of $\sim$0.02
mJy beam$^{-1}$ at 45\,GHz. To extract the flux densities and
uncertainties from the radio maps, we used \texttt{`imfit'}, part of
the CASA \texttt{`viewer'}, which fits two-dimensional Gaussians to an
intensity distribution on a region selected interactively on the
map. To possibly increase the chance of detecting the core and
  extended emission, we also reduced the map resolution by using
  different values of the \verb'UVTAPER' parameter in the \verb'IMAGR'
  task, ranging between 600\, k$\lambda$ to 250 k$\lambda$ for all the
  sources. This parameter specifies the width of the Gaussian
function in the $uv$-plane to down-weight long-baseline data
points. Smaller units of this parameter correspond to a shallower
angular resolution, i.e. a larger beam size, up to a factor $\sim$10
larger than those reached in full resolution (i.e. $\sim$0.3-0.5
arcsec).

\begin{table*}
\centering
\caption{VLA peak flux densities in mJy of the 15 observed PG RQQs from 1.4
  to 15 GHz taken from VLA archive and literature.}
\begin{tabular}{l|ccc|cccc}
\hline
                &  \multicolumn{3}{c|}{Array-A VLA flux densities} & \multicolumn{4}{c}{B-C-D VLA flux densities} \\
 Frequency:     &   1.4GHz   &  5 GHz  &  8.5 GHz  &  1.4GHz   &  5 GHz  &  8.5 GHz  &  15 GHz \\
Resolution:   &  1\farcs3  &  0\farcs33  &  0\farcs20  &    &           &             &\\
\hline
 Object            &   $F_\nu$       &  $F_\nu$    &   $F_\nu$    &   $F_\nu$       &  $F_\nu$    &   $F_\nu$  &   $F_\nu$  \\
\hline
\multirow{2}{1.8cm}{\centering PG~0003+199} &  \multirow{2}{1.8cm}{\centering 6.74$\pm$0.11$^{a}$}   &    3.03$\pm$0.10$^{b}$   &   \multirow{2}{1.8cm}{\centering 2.05$\pm$0.1$^{c}$}  &  \multirow{2}{1.8cm}{\centering 7.3$\pm$0.5$^{f}$}  &    3.92$\pm$0.1$^{b}$  &    \multirow{2}{1.8cm}{\centering 2.23$\pm$0.25$^{g}$} \\
            &                      &    3.49$\pm$0.05$^{e}$    &          &        &    3.58$\pm$0.1$^{e}$                 \\
PG~0026+129 &  2.4$\pm$0.7$^{d}$   &    0.20$\pm$0.06$^{b}$    &  0.17$\pm$0.05$^{e}$  &  7.1$\pm$0.5$^{f}$  & 5.10$\pm$0.1$^{b}$ &  2.37$\pm$0.06$^{e}$ &  \\
PG~0049+171 &  0.44$\pm$0.12$^{a}$ &    0.66$\pm$0.10$^{b}$    &                  &  $<$2.5$^{f}$   & 0.64$\pm$0.10$^{b}$ &  0.66$\pm$0.10$^{g}$   &            \\
\multirow{2}{1.8cm}{\centering PG~0050+124} &  \multirow{2}{1.8cm}{\centering 5.1$\pm$0.4$^{d}$}   &   1.80$\pm$0.10$^{b}$    &   \multirow{2}{1.8cm}{\centering 0.9$\pm$0.2$^{d}$}  &  \multirow{2}{1.8cm}{\centering 6.22$\pm$0.35$^{h}$}  &  \multirow{2}{1.8cm}{\centering 2.60$\pm$0.11$^{b}$} &        &   \multirow{2}{1.8cm}{\centering 1.06$\pm$0.32$^{h}$}  \\
            &                     &    2.4$\pm$0.3$^{d}$      &                      \\
\multirow{2}{1.8cm}{\centering PG~0052+251} &  \multirow{2}{1.8cm}{\centering 1.1$\pm$0.4$^{d}$}  &    0.42$\pm$0.10$^{b}$   &   \multirow{2}{1.8cm}{\centering 0.7$\pm$0.1$^{d}$} &  \multirow{2}{1.8cm}{\centering 1.59$\pm$0.14$^{a}$}  &   0.74$\pm$0.10$^{b}$ &     &    \\
            &                     &    0.58$\pm$0.03$^{e}$    &       &    &        0.61$\pm$0.03$^{e}$  &   &                  \\
PG~0157+001 &                     &    5.58$\pm$0.06$^{b}$    &  2.98$\pm$0.03$^{e}$  &  22.5$\pm$0.8$^{h}$ &  7.88$\pm$0.26$^{h}$ &  4.23$\pm$0.03$^{e}$ & 2.04$\pm$0.32$^{h}$ \\
PG~0844+349 &                     &    $<$0.25$^{b}$          &                     & $<$0.38$^{i}$    &  0.31$\pm$0.06$^{b}$  &      &     \\
PG~1001+054 &                     &    $<$0.25$^{b}$          &                  &    $<$2.5$^{f}$   &    0.80$\pm$0.1$^{b}$  &  0.64$\pm$0.3$^{a}$ & \\
PG~1048+342 &                     &                          &                   &   $<$0.42$^{i}$  &  $<$0.19$^{b}$    &   $<$0.12$^{a}$ &  \\
PG~1049-005 &                     &   0.22$\pm$0.03$^{e}$    &                   &  0.82$\pm$14$^{i}$ & 0.48$\pm$0.03$^{b}$  &  0.37$\pm$0.07$^{a}$   &  \\
PG~1114+445 &                     &   0.20$\pm$0.06$^{b}$    &                   &  0.71$\pm$0.12$^{i}$ & 0.22$\pm$0.06$^{b}$ &        &    \\
\multirow{3}{1.8cm}{\centering PG~1149-110} &  \multirow{3}{1.8cm}{\centering 3.1$\pm$0.8$^{d}$}  &   1.20$\pm$0.20$^{d}$    & 1.3$\pm$0.1$^{d}$  &  \multirow{3}{1.8cm}{\centering 9.7$\pm$0.6$^{f}$} &  2.60$\pm$0.1$^{b}$  &    \multirow{3}{1.8cm}{\centering 0.54$\pm$0.02$^{a}$} & \\
            &                     &   1.08$\pm$0.06$^{e}$    & 0.99$\pm$0.04$^{a}$   &     &      1.94$\pm$0.1$^{e}$  &   \\
            &                     &   1.0$\pm$0.1$^{b}$      &                       \\ 
PG~1229+204 &                     &   0.30$\pm$0.06$^{b}$    &              &   2.8$\pm$0.4$^{f}$   & 0.67$\pm$0.1$^{b}$   &   &     \\ 
PG~1259+593 &                     &   $<$0.25$^{b}$          &             &  $<$0.45$^{i}$   &  $<$0.3$^{a}$     &    &    \\ 
PG~1310-108 &                     &   $<$0.25$^{b}$          &             & $<$2.5$^{f}$  &  0.26$\pm$0.06$^{b}$ &   $<$0.15$^{a}$           \\ 
\hline
\end{tabular}
\label{coreflux}
\begin{flushleft}

Upper limits are evaluated at 3$\sigma$ level. References: $a$ map
obtained from the image VLA archive, $b$ \citet{kellermann89}, $c$
\citet{kukula95}; $d$ \citet{kukula98}; $e$ \citet{leipski06}; $f$
\citet{NVSS}; $g$ \citet{barvainis05}; $h$ \citet{barvainis89}; $i$
\citet{FIRST}.
\end{flushleft}
\end{table*}

\section{Results}
\label{results}

\begin{figure*}
  \vspace{-0.3cm}
  \centerline{
 \includegraphics[angle=90,width=0.8\columnwidth]{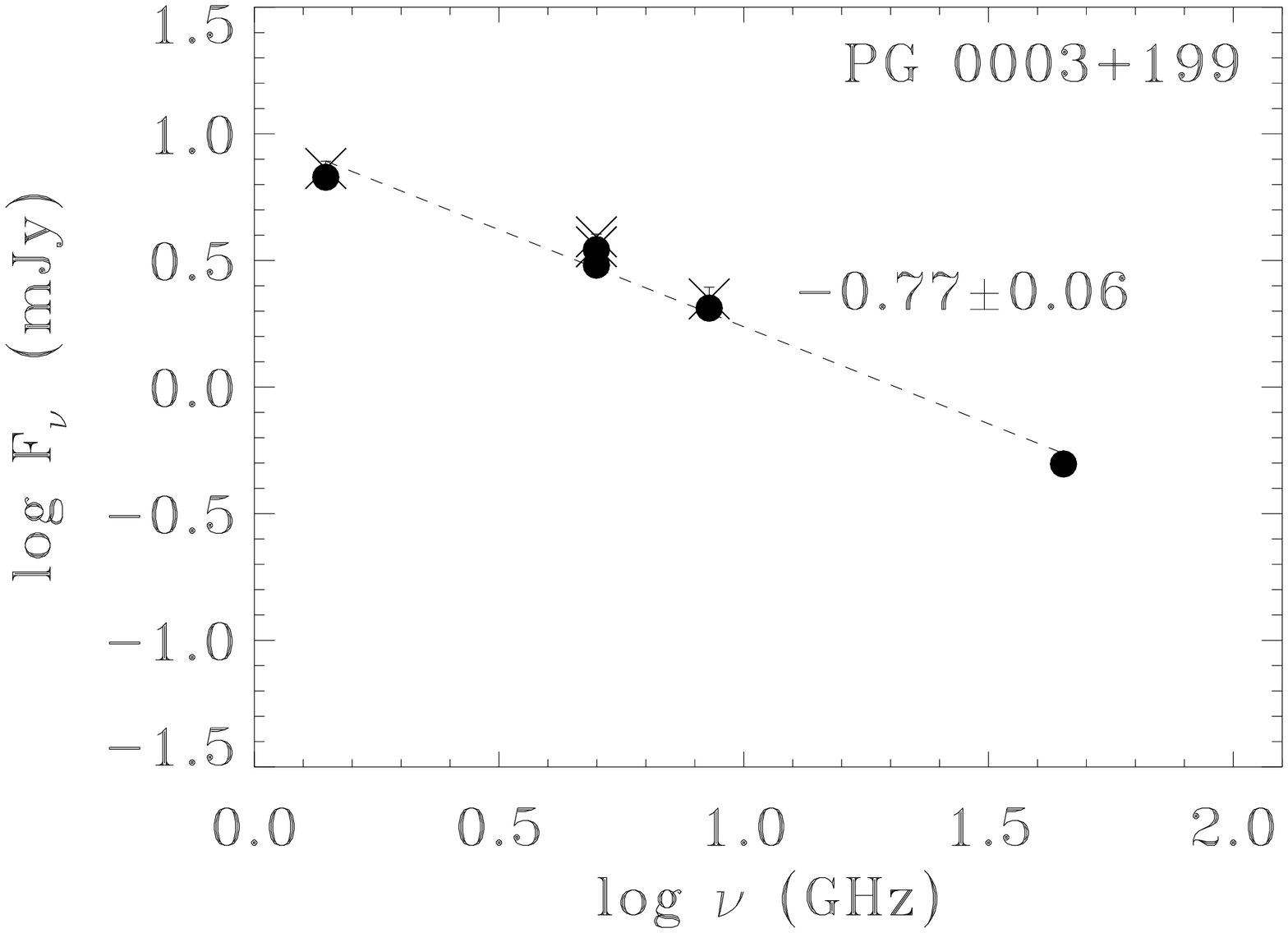}
  \includegraphics[angle=90,width=0.8\columnwidth]{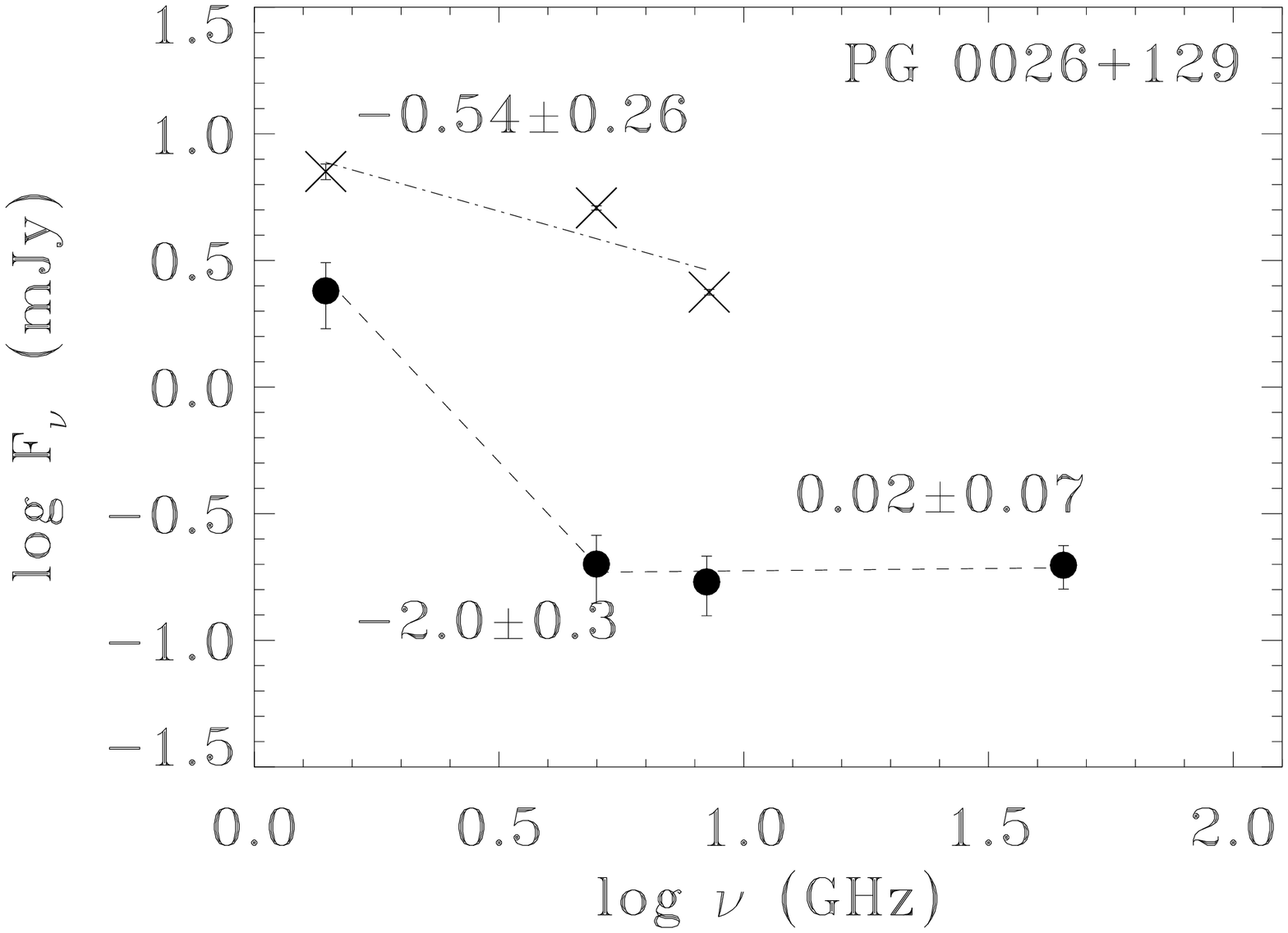}
  \includegraphics[angle=90,width=0.8\columnwidth]{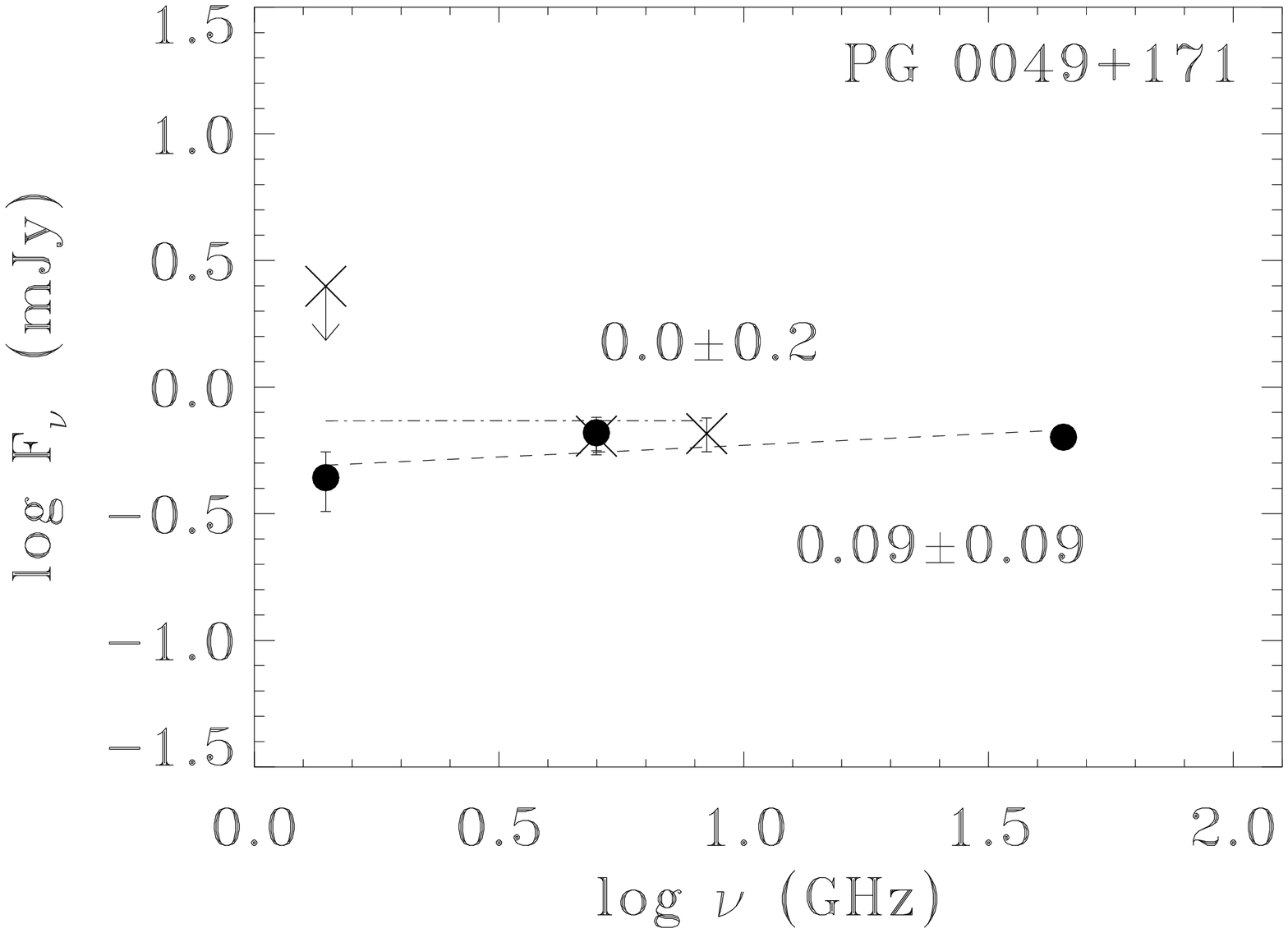}}
  \vspace{-0.5cm}
\centerline{
 \includegraphics[angle=90,width=0.8\columnwidth]{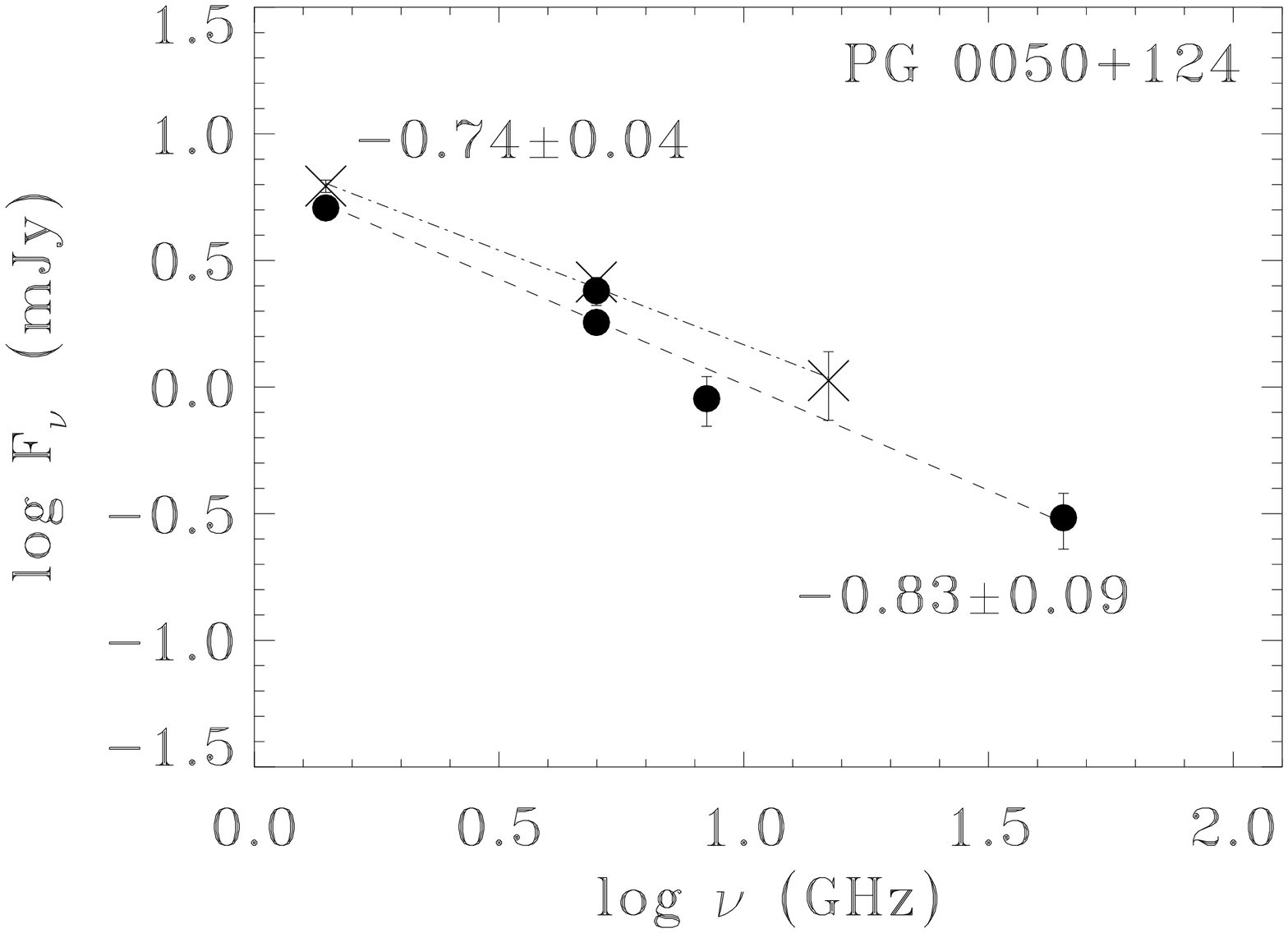}
  \includegraphics[angle=90,width=0.8\columnwidth]{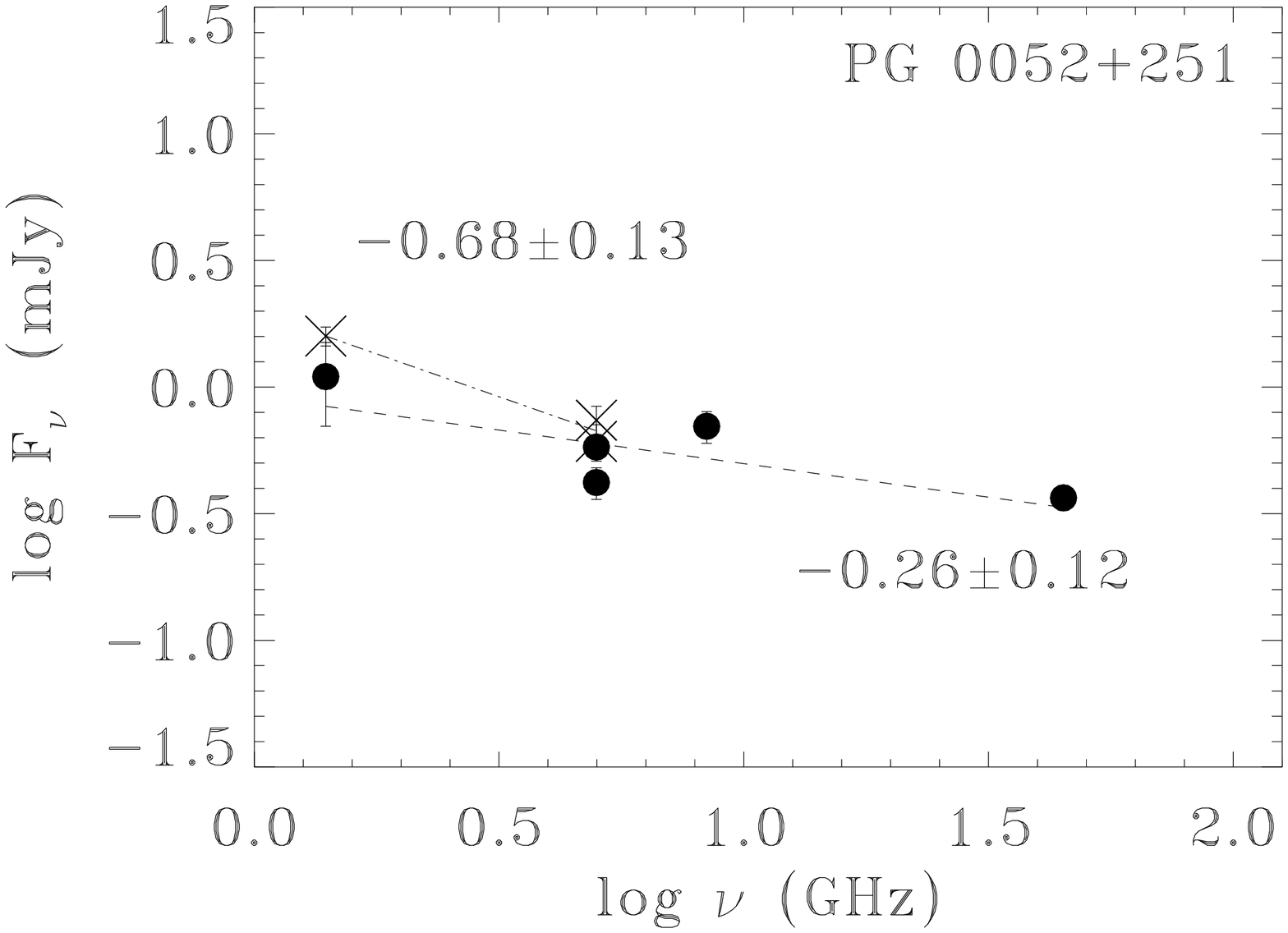}
  \includegraphics[angle=90,width=0.8\columnwidth]{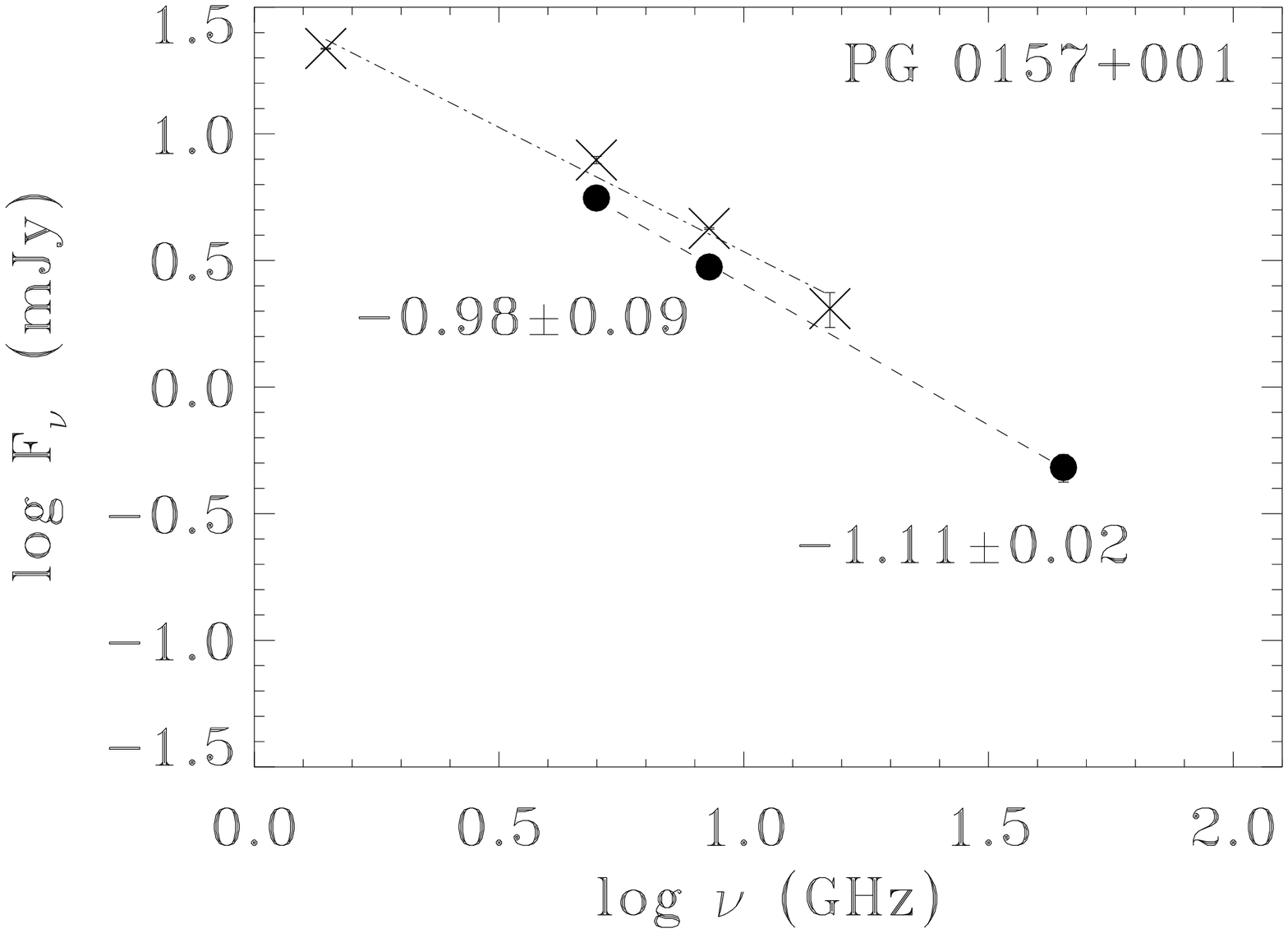}}
  \vspace{-0.5cm}

\centerline{
 \includegraphics[angle=90,width=0.8\columnwidth]{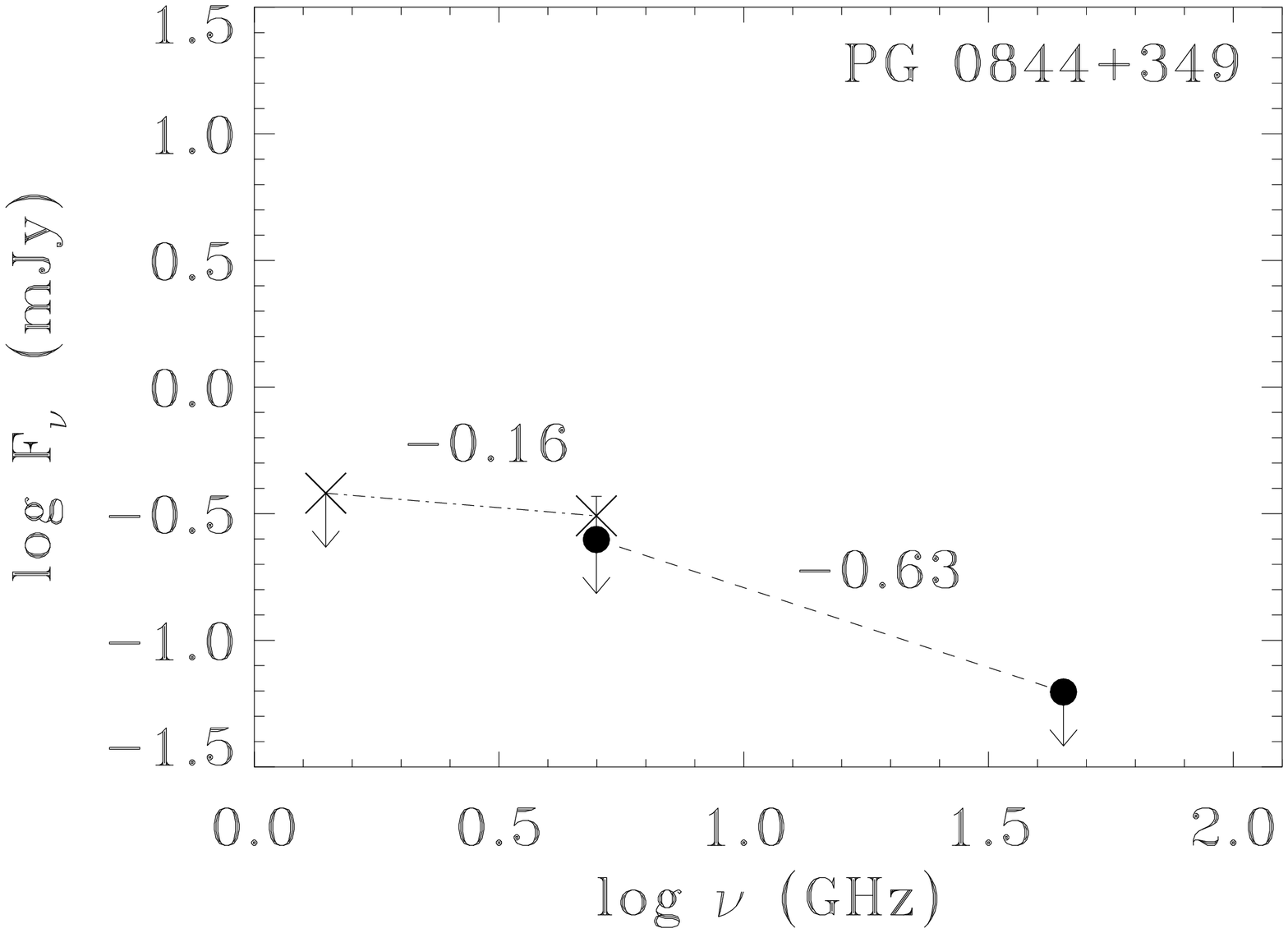}
  \includegraphics[angle=90,width=0.8\columnwidth]{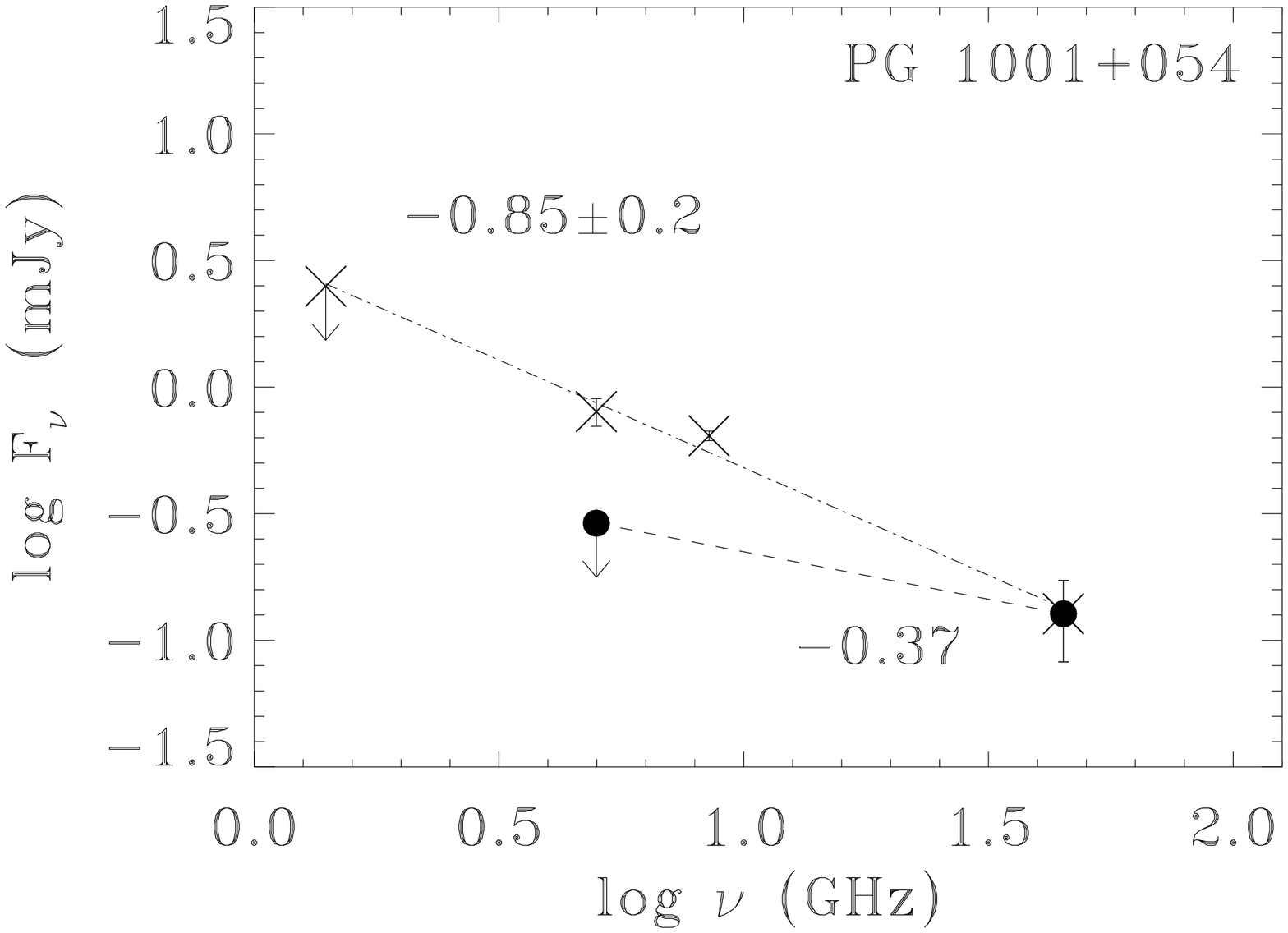}
  \includegraphics[angle=90,width=0.8\columnwidth]{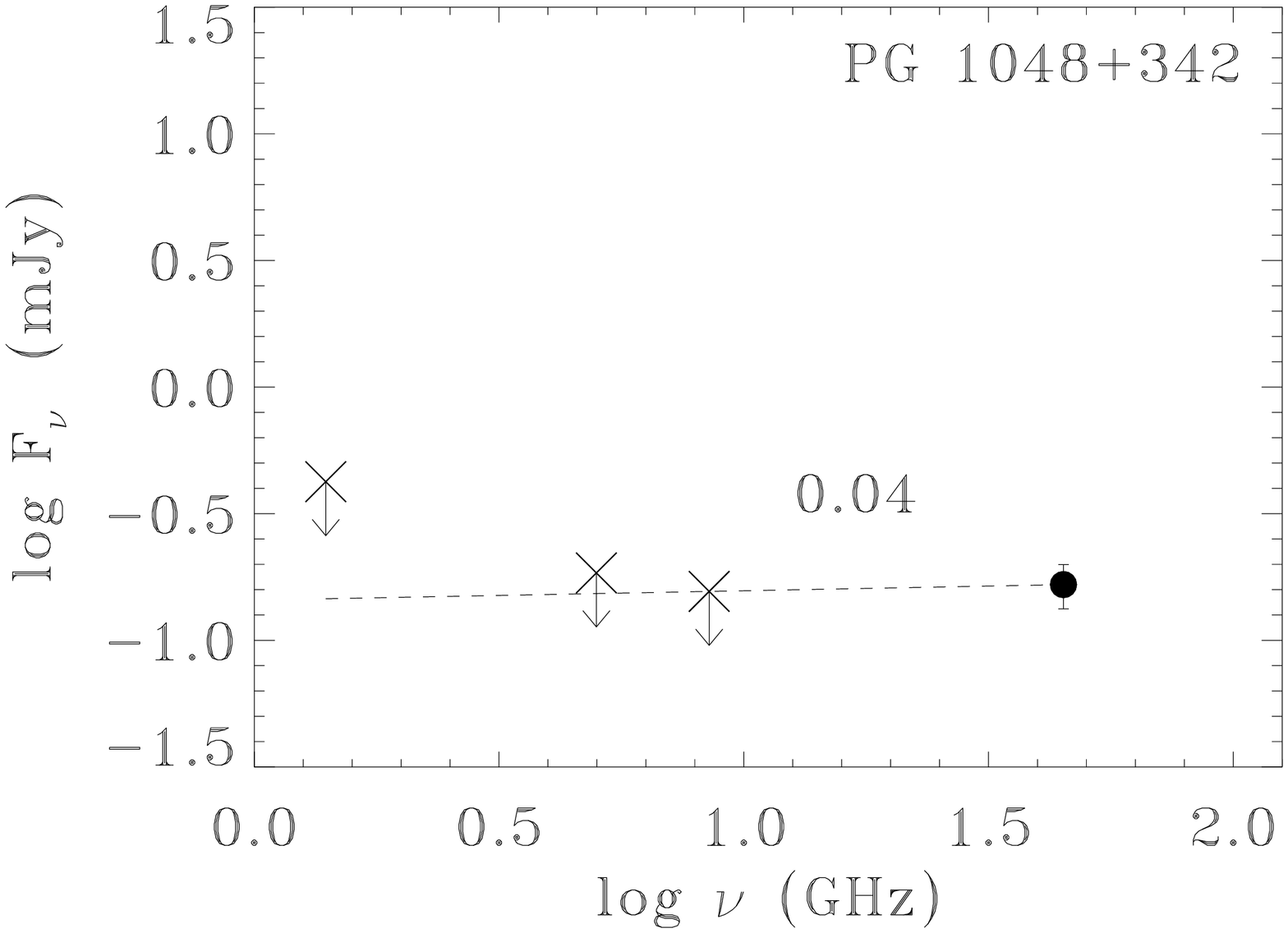}}
  \vspace{-0.5cm}

\centerline{
 \includegraphics[angle=90,width=0.8\columnwidth]{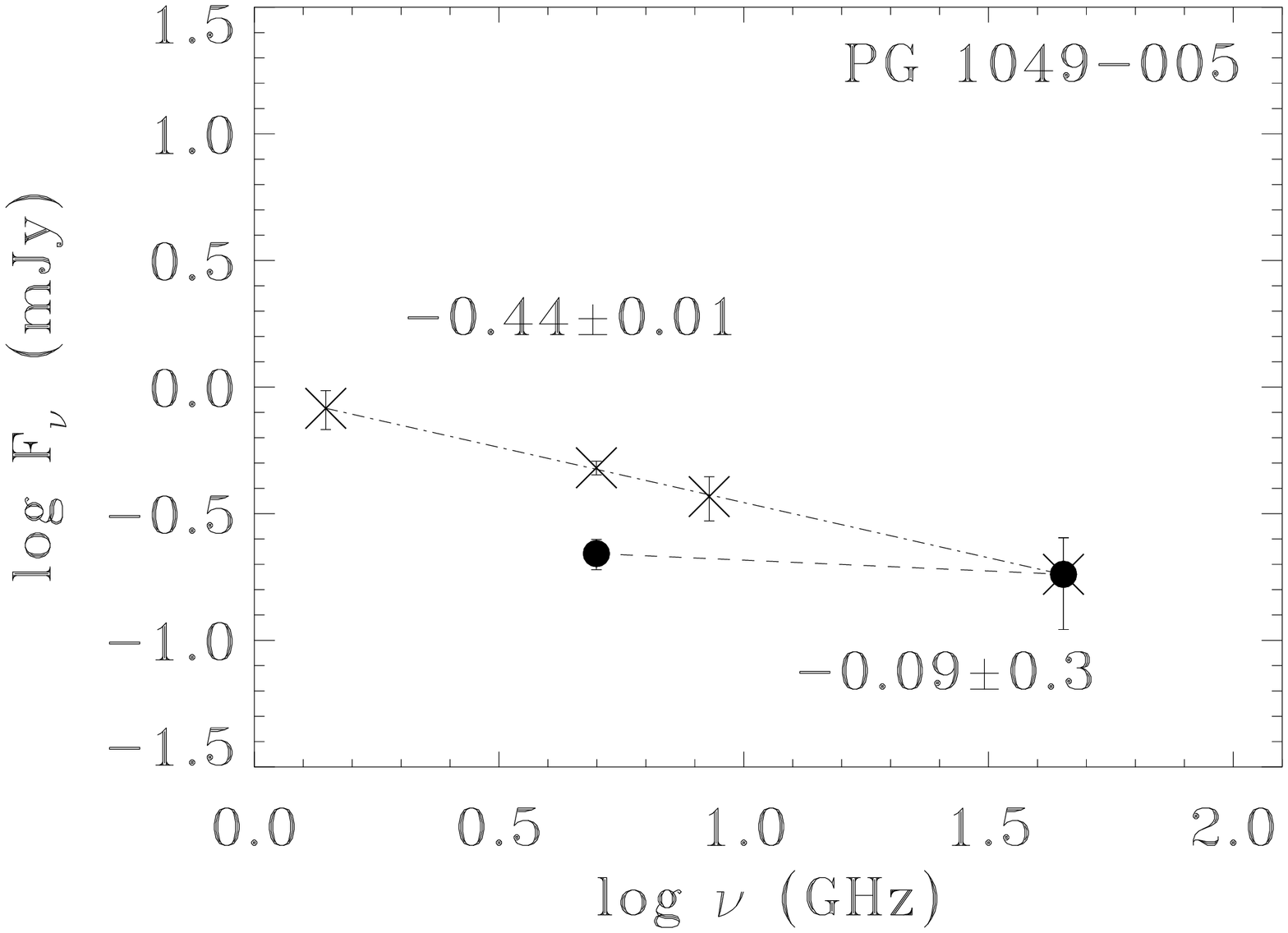}
  \includegraphics[angle=90,width=0.8\columnwidth]{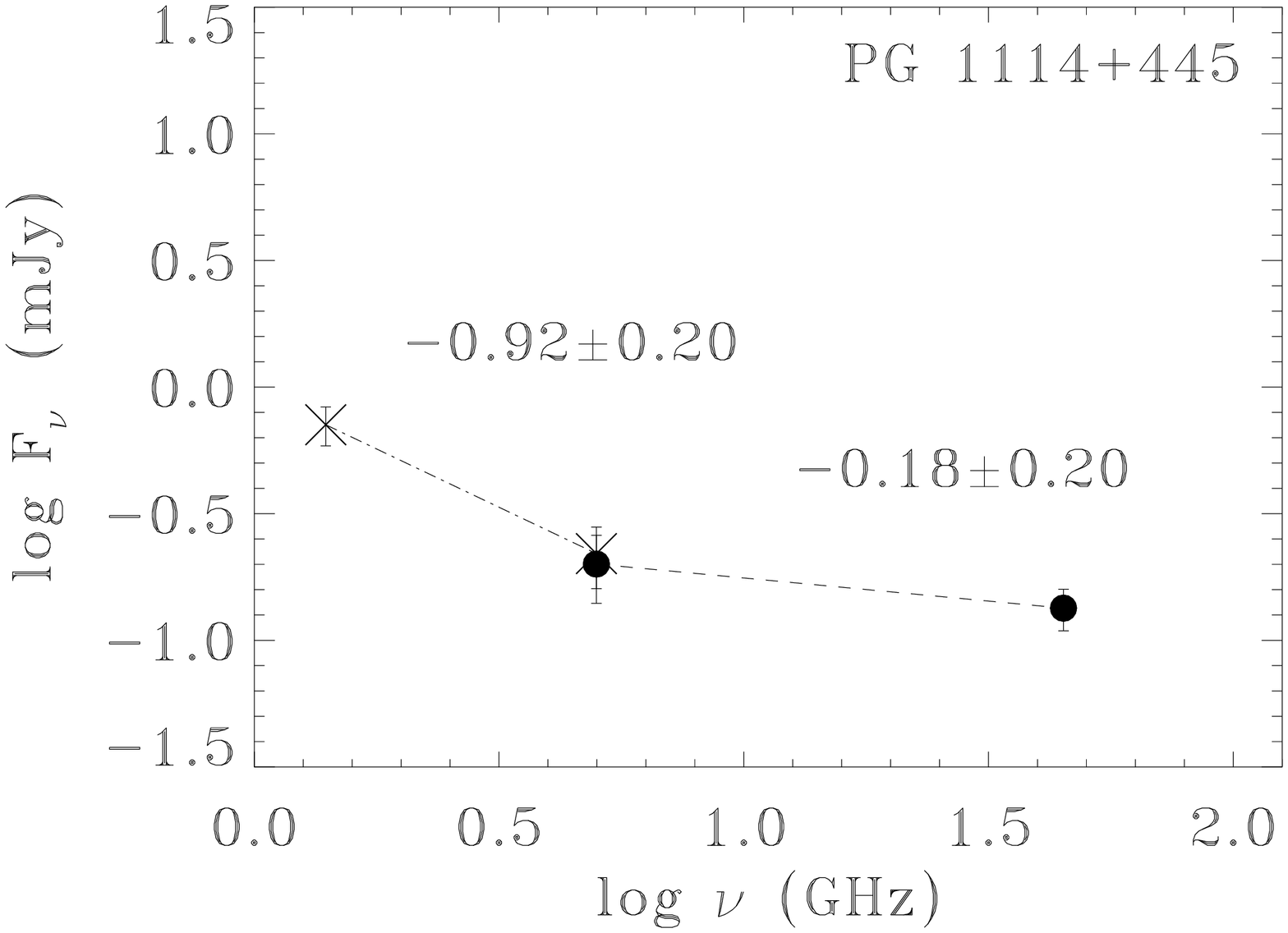}
  \includegraphics[angle=90,width=0.8\columnwidth]{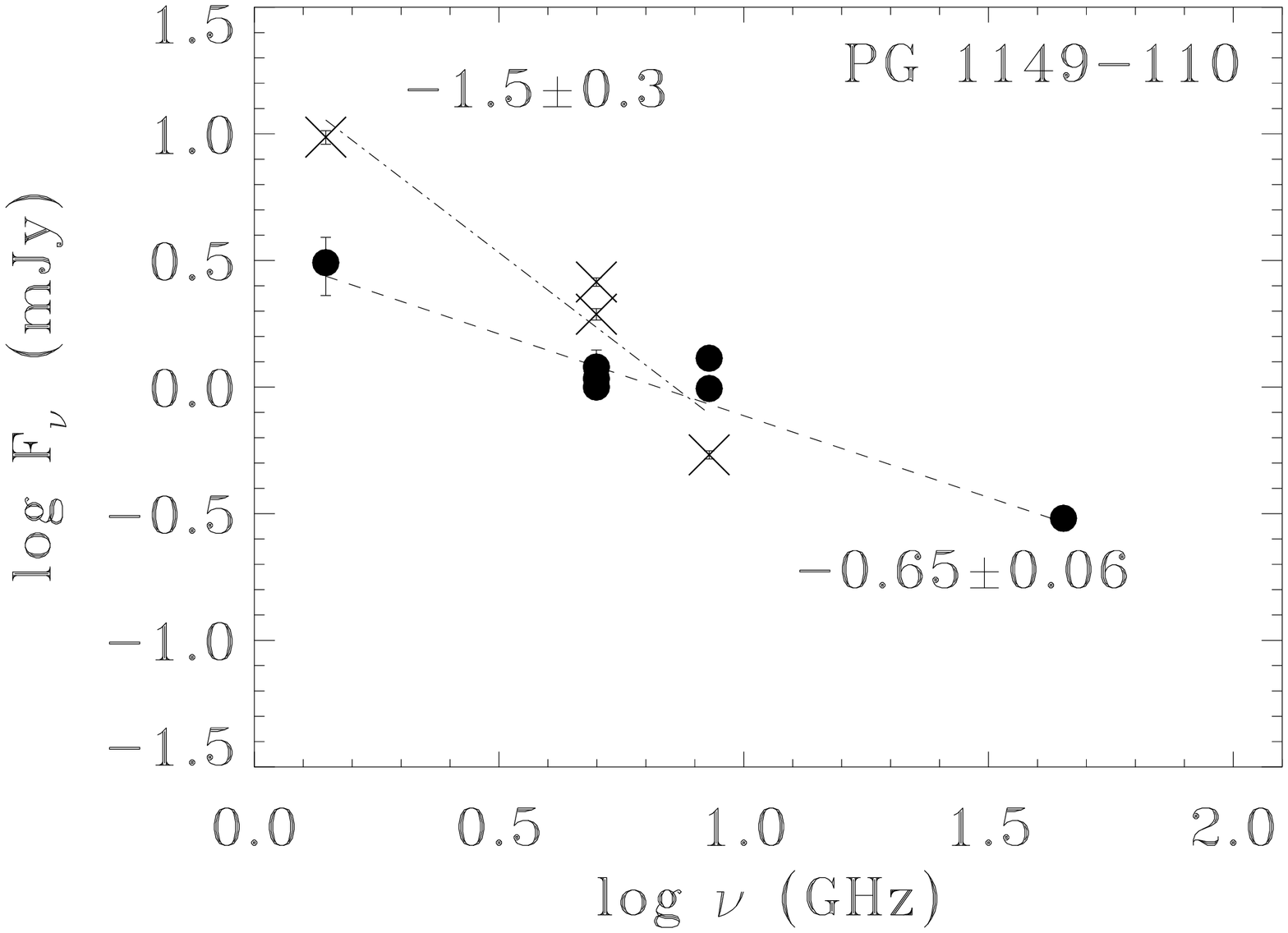}}
  \vspace{-0.5cm}

\centerline{
 \includegraphics[angle=90,width=0.8\columnwidth]{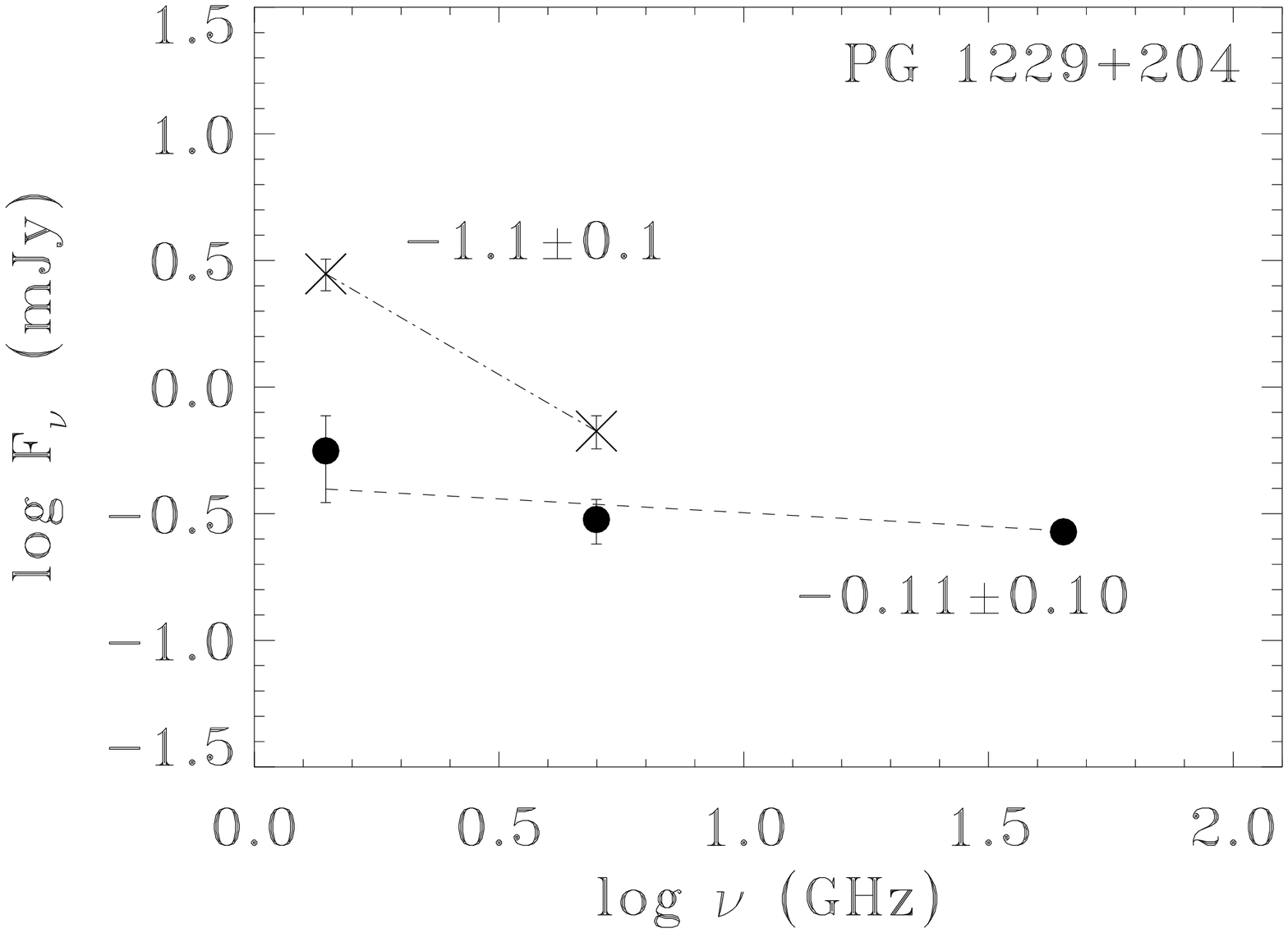}
  \includegraphics[angle=90,width=0.8\columnwidth]{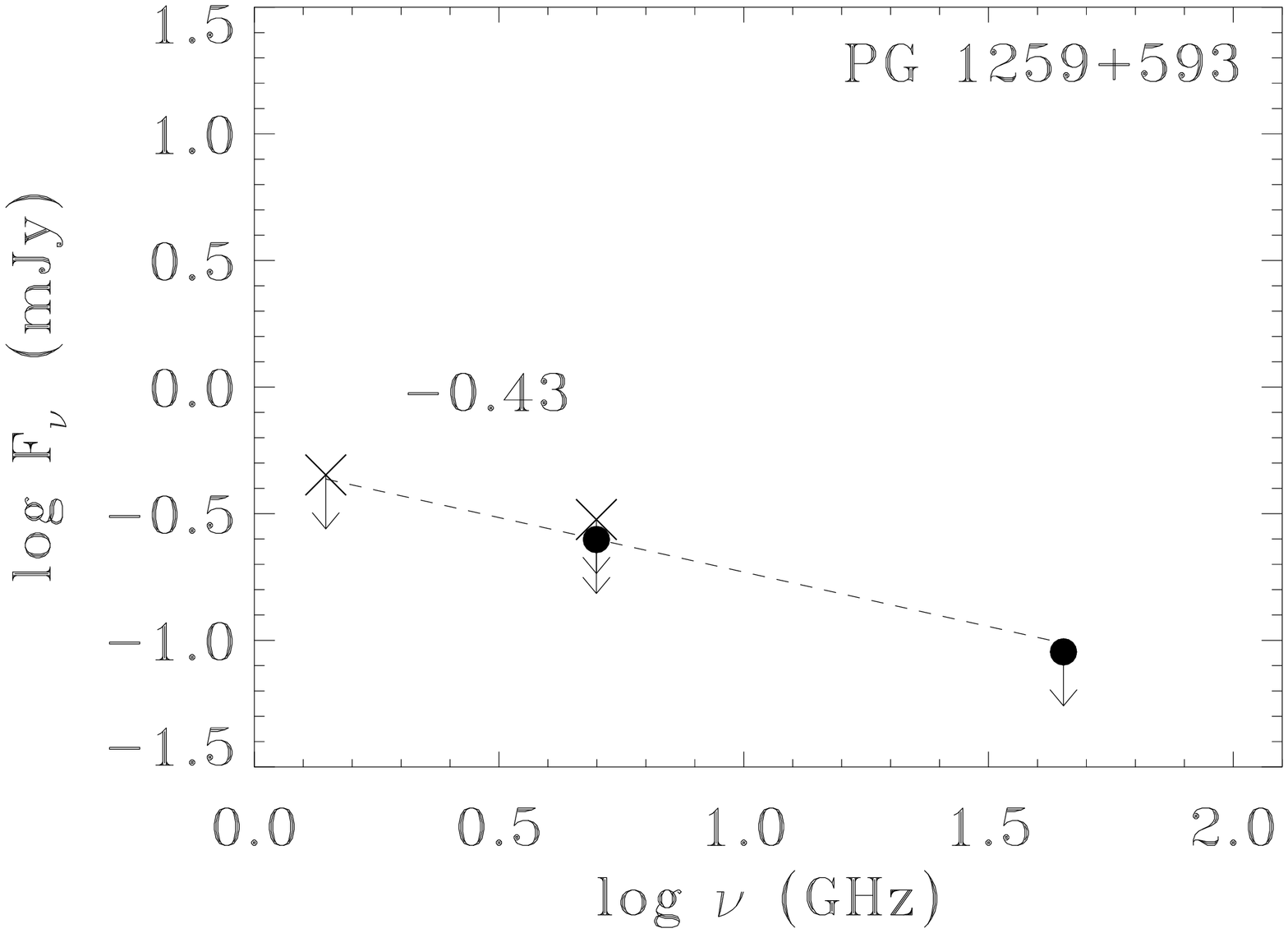}
\includegraphics[angle=90,width=0.8\columnwidth]{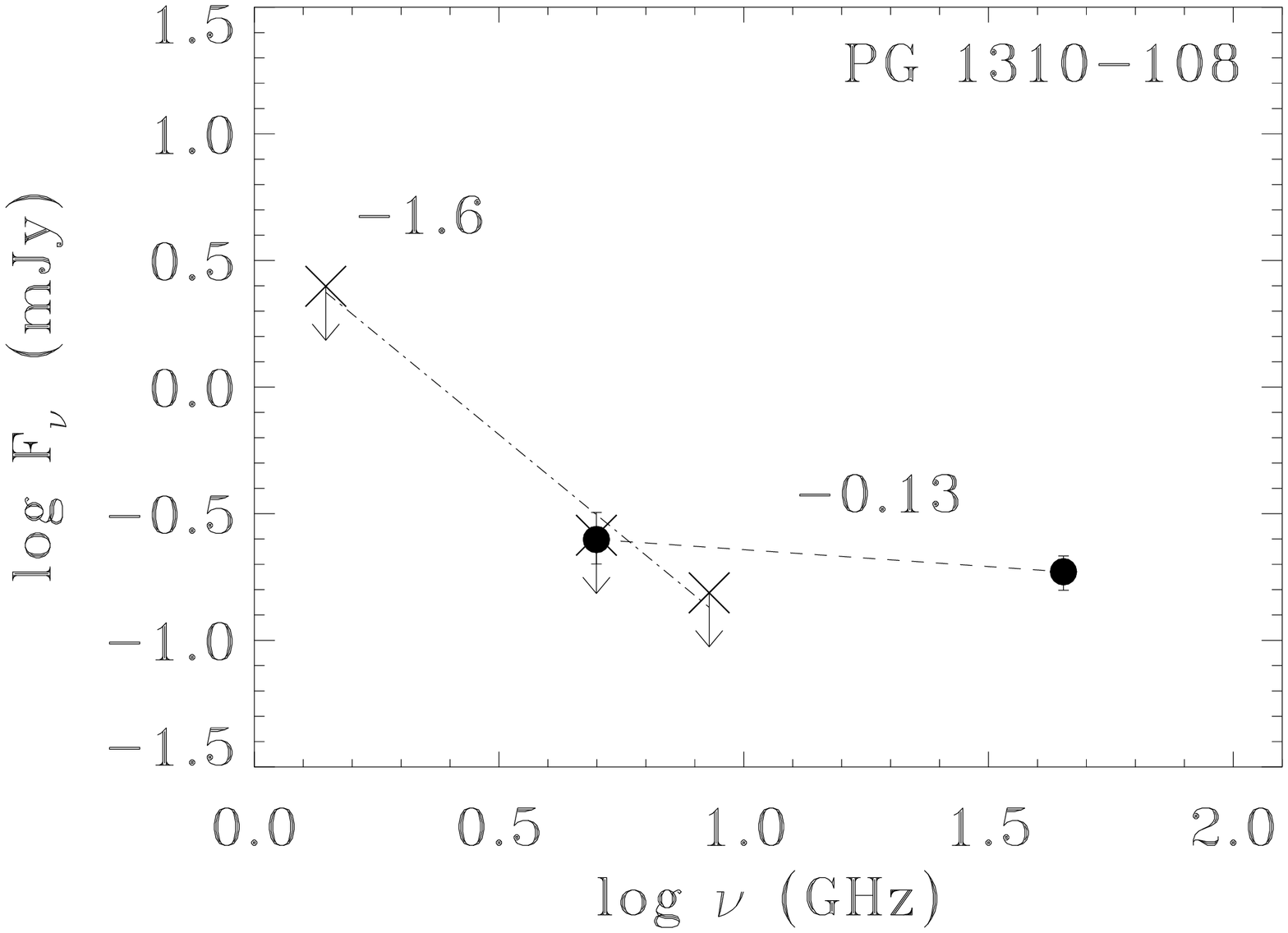}}
\vspace{-0.3cm}
\label{sed}
\caption{A compilation of the overall 1.4--45~GHz SED of the 15 PG RQQs
  presented in this work based on VLA data. Filled points mark the
  highest resolution, A array, data, and the crosses measurements at
  lower resolutions (B-C-D arrays). The dot-dashed line shows the
  spectral slope for the VLA low-resolution data, while the dashed
  line represents the spectral properties of the unresolved core
  component at VLA high resolution. The corresponding values reported
  on the plot are the spectral indices derived by fitting the censored data.} 
\end{figure*}

\begin{figure}
  \includegraphics[width=0.7\columnwidth,angle=90]{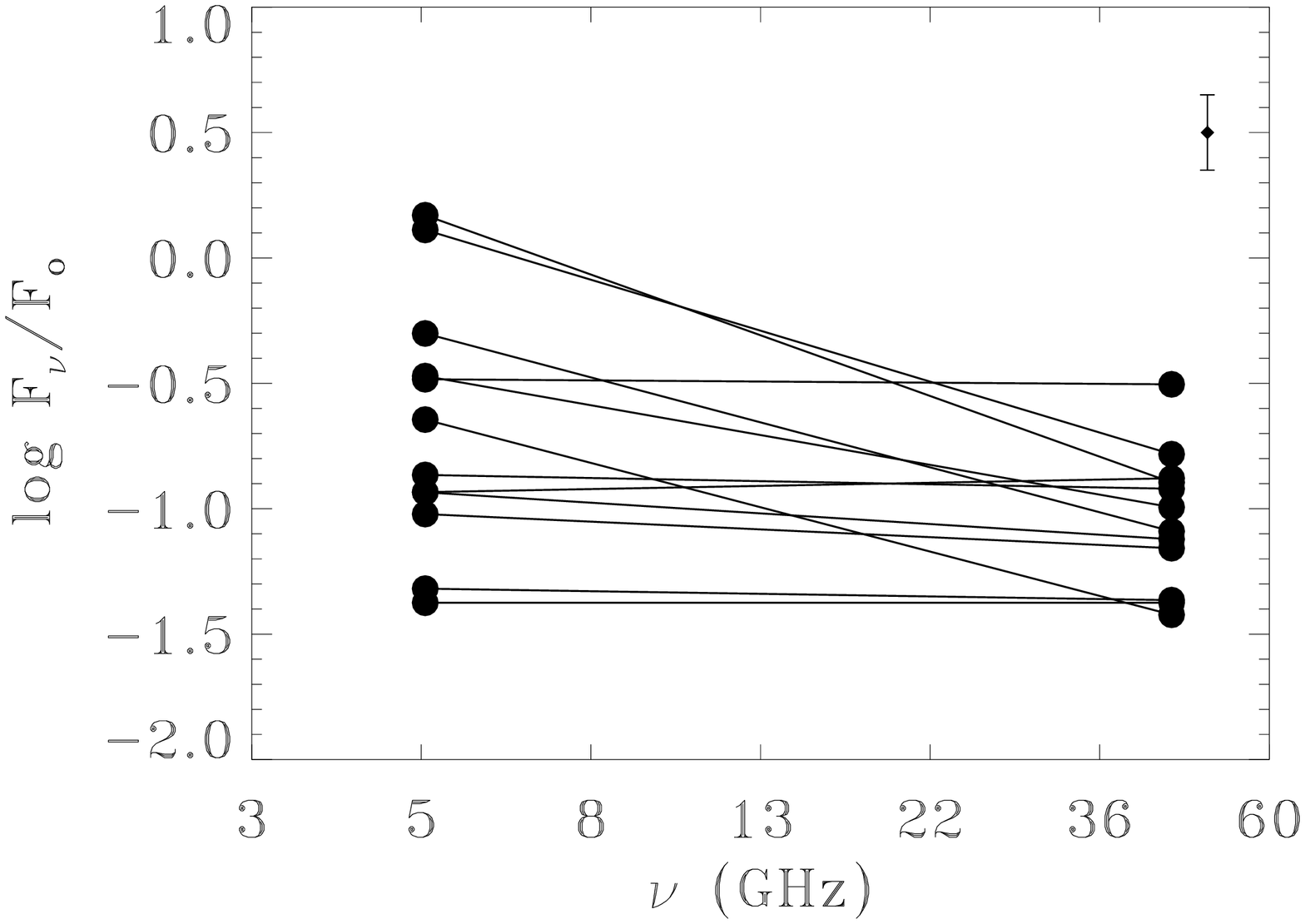}
\label{spectrum}
\caption{Radio flux densities at 5 and 45 GHz (in logarithmic scale)
  normalised by the optical 4400\AA\, flux density (i.e ratio between
  the radio and optical flux densities) for the RQQs which are
  detected in Q-band. At the top-right corner, we draw the typical
    error bar on the y-axis, $\sim$0.2-0.3. The plot shows that
  RQQs tend to have steeper radio spectra because of brighter 5~GHz
  emission relative to the optical, rather than dimmer 45~GHz
  emission. }
\end{figure}

In the full-resolution maps we detect radio emission from the central
unresolved core in 10 sources with a flux density of the order of
$\sim$0.1 mJy beam$^{-1}$, reaching a S/N $\sim$ 3-5 (see
Table~\ref{lum}). Reducing the angular resolution by up to a factor
$\sim$10 with \verb'UVTAPER' allows us to detect three additional
objects, PG~0050+124,PG~1001+054, and PG~1049-005. However, two
objects, PG~0844+349 and PG~1259+593, remain undetected. In conclusion,
we detect a 45-GHz radio core in 13 out of 15 sources (86 $\pm$ 7 per
cent of our sample).

Figure~\ref{maps} presents the full-resolution radio images at 45\,GHz
of our sample. For the sources not detected in full resolution, we
display the $uv$-tapered images.  Most of the targets appear
unresolved or slightly resolved at the full resolution on 50-100~mas
(Table~\ref{tab_cont}) which corresponds to a few hundred pc
(Table~\ref{lum}). The ratio between total and peak flux densities for
these sources is close to unity, suggesting their structure is
unresolved. Four sources (PG~0049+171, PG~0050+124, PG~0157+001 and
PG~1114+445) show a marginal elongation of the radio core
emission. The $uv$-tapered maps do not show any significant
  evidence of extended emission in any of the targets of the sample.

\begin{figure*}
  \includegraphics[width=1.6\columnwidth,angle=90]{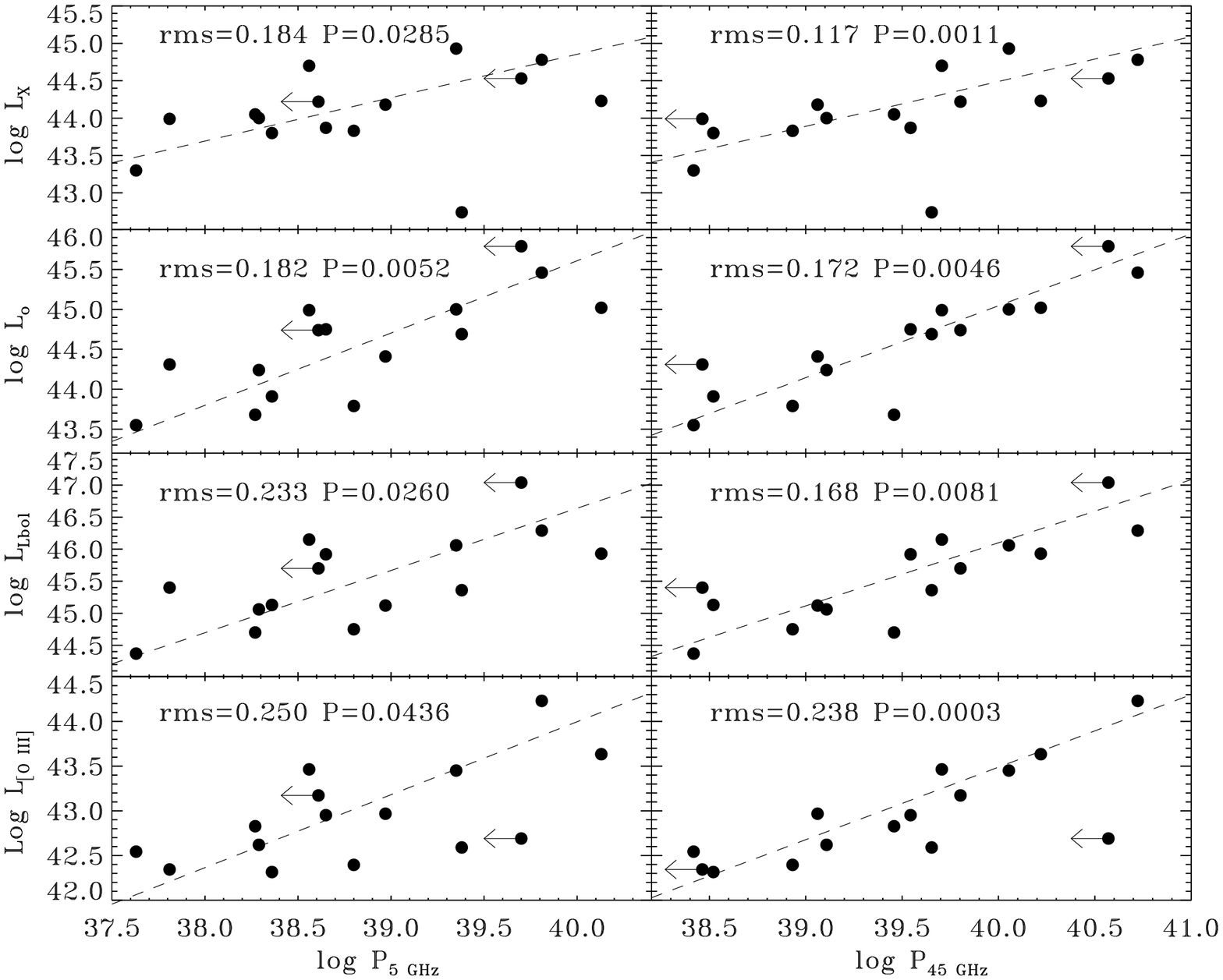}
  \vspace{-0.5cm}
\caption{The relation between the radio core luminosities at 5 and 45
  GHz (erg s$^{-1}$) and the X-ray (2-10 keV), optical, bolometric and
  [O~III] line luminosities (erg s$^{-1}$) for the 15 PG RQQs. The
  value in the upper part of each panel expresses the rms scatter of
  the relation (assuming the upper-limits as detections) and the
  generalised Kendall's $\tau$ test probability (P) of a fortuitous
  correlation. Note that the relations with P$_{45~{\rm GHz}}$ are
  generally tighter, and the tightest relation is with L$_{\rm
    X}$. The outlier PG~1001+054 with L$_{\rm X} < 10^{43}$ erg
  s$^{-1}$ is known to be significantly X-ray absorbed
  \citep{schartel05}, and was not considered in fitting the X-ray--radio
  relations.}
\label{multi_5_45}
\end{figure*}

\begin{table}
\begin{center}
  \caption[Properties of the RQQ sample.]{Radio spectral indices of the PG 15 RQQs}
\begin{tabular}{l|cccc}
  \hline
  name        &  1 comp        &   2 comp        & S/F &  $\alpha_{5-45\rm{GHz}}$ \\
  \hline
  PG~0003+199 & -0.77$\pm$0.06 &                 & $S$   &   -0.83 \\
  PG~0026+129 & -0.54$\pm$0.26 &  0.02$\pm$0.07  & $S+F$   &    0.20  \\
  PG~0049+171 & 0.0$\pm$0.2    &  0.09$\pm$0.09  & $F$ &    0.25  \\
  PG~0050+124 & -0.74$\pm$0.04 &  -0.82$\pm$0.09 & $S$   &   -0.90  \\
  PG~0052+251 & -0.68$\pm$0.13 & -0.26$\pm$0.12  & $S+F$ &   -0.26  \\
  PG~0157+001 & -0.98$\pm$0.09 & -1.11$\pm$0.02  & $S$   &   -0.91  \\
  PG~0844+349 & -0.16          &  -0.63          & $S$   &   $<$-0.32  \\
  PG~1001+054 & -0.85$\pm$0.20 & -0.37           & $S+F$ &   -0.71  \\
  PG~1048+342 & 0.04           &                 & $F$   &    $>$0.25  \\
  PG~1049-005 & -0.44$\pm$0.01   &  -0.09$\pm$0.30 & $F$   &  -0.04   \\
  PG~1114+445 & -0.92$\pm$0.20 & -0.18$\pm$0.20  & $S+F$ &  -0.06  \\
  PG~1149-110 & -1.5$\pm$0.3   & -0.65$\pm$0.06  & $S$   &   -0.86  \\
  PG~1229+204 & -1.1$\pm$0.1   & -0.11$\pm$0.10  & $S+F$ &   -0.14   \\
  PG~1259+593 &  -0.43         &                 & ?   &         \\
  PG~1310-108 &  -1.6          & -0.13           & $F$ &   -0.17  \\
\hline
\hline                                          
\end{tabular}                                   
\label{tab_alfa}
\begin{flushleft}
  Column description: (1) PG source name; (2)-(3) spectral slopes of
  the first and second radio components, estimated respectively
  typically at lower frequencies ($\lesssim$5 GHz) and lower
  resolution (VLA B-C-D array), and higher radio frequencies
  ($\gtrsim$5 GHz) and higher resolution (VLA A array) (see Fig.~3);
  (4) general radio SED dominance: $S$ for steep-spectrum component,
  $F$ for flat-spectrum component; $S+F$ combination of a steep- and
  flat-spectrum component; (5) spectral indices between 5 and 45 GHz,
  with typical errors of 30 per cent.
\end{flushleft}
\end{center}
\end{table}

With the final goal of deriving the radio cm-wavelengths SEDs, we
search in the literature and VLA image
archive\footnote{https://archive.nrao.edu/archive/archiveimage.html}
for radio data below 45~GHz, both at the highest resolution with A
array, and at the lower resolution with B-C-D arrays (see
Table~\ref{coreflux} and references therein). Figure~\ref{sed}
displays the radio SEDs of the 15 objects in the frequency range 1.4
-- 45 GHz. Filled points mark the highest resolution A array data, and
the crosses represent the measurements at lower resolutions. There is
a rich variety of spectral shapes ($F_{\nu} \sim \nu^{\alpha}$) across
the 1--45 GHz band (Tab.~\ref{tab_alfa}). We define the
  broad-band spectral class of the sources, based on the both the
  A-array and the lower resolution VLA data available for the
  targets:

  \begin{itemize}
  \item {\it steep (S)}: if the radio SED is characterised by a steep
    spectral index ($\alpha <$ -0.5) in both high and low resolution
    data: PG~0003+199, PG~0050+124, PG~0157+001, PG~1149-110. We also
    include PG~0844+349 into this category because with only one
    detection (at low resolution at 5 GHz) the estimated spectral
    slope can be considered as upper limit.

  \item {\it flat or slightly inverted (F)}: if the radio SED is
    characterised by a flat spectral index (0.1 $> \alpha >$ -0.5) in
    both high and low resolution data (or only in high resolution if
    the low resolution data do not provide detections): PG~0049+171,
    PG~1048+342, PG~1049-005, and PG~1310-108.

  \item {\it a broken
power-law (S+F)}: the combination of steeper lower-frequency spectrum
    and a flatter higher-frequency spectrum at either low or high resolution data: PG~0026+129, PG~0052+251, PG~1001+054, PG~1114+445, and  PG~1229+204.

\end{itemize}
    
  We abstain from the spectral classification of PG~1259+593 because
  of the lack of detections from the available radio data.
Note that these radio SEDs do not account for radio core variability,
which is known to be present by as much as a factor of 2-3
\citep{falcke01,barvainis05,mundell09,baldi15,behar20,nyland20}, nor the
different angular resolutions ranging from 1.3 to 0.40 arcsec with
A-array observations in different bands.

The radio SEDs presented in Fig.~\ref{sed} suggest that the typical
radio spectrum of PG RQQs consists of a combination of at least two
components\footnote{The spectral slopes are obtained by fitting the
censored radio data in Fig.~\ref{sed}. In case of the presence of
non-detections, the spectral slope are rough estimates, measured by
assuming upper limits as detections for simplicity.}, a
steep-spectrum component which dominates at low frequencies and low
resolution, and a flat-spectrum component (typically unresolved) which
emerges at high resolution and at high frequencies
(Tab.~\ref{tab_alfa}). At least for 9 sources a flat-spectrum radio
component emerges at high radio frequencies. Is this because one
component is brighter, or because the other component is weaker,
relative to the emission in other bands?

Figure~4 shows the 5--45 GHz spectra of the detected PG
RQQs normalised by the flux densities in the optical band from
\citet{kellermann89} at 4400 \AA\, (a proxy of radio loudness). The plot
shows that, the spectral steepness of the RQQs is due to the
brighter 5-GHz core component, rather than a dimmer 45-GHz
component. In turn, the flattening of the RQQ spectra is the result of
the weakening of the steep-spectrum component.  This results would suggest
that steep-spectrum objects have an additional mechanism which produces
optically thin emission, which adds to the flat-spectrum optically thick
emission that most of objects have.

\subsection{Multi-band correlations}
\label{multicorr}

As higher-energy photons are expected to originate from inner regions
of the nucleus (accretion disc, jet base, etc), early studies revealed
that radio properties in RQQs are linked to the
accretion/ejection activity, derived from high-energy bands (e.g.,
optical, X-ray, \citealt{boroson92,laor08}). In addition, it has been
recently found the radio spectral slopes of RQQs can also depend on
several AGN parameters, which in turn are linearly combined in the EV1
diagram (see \citealt{laor19} and references therein).

To derive possible meaningful correlations between the 45-GHz emission
with that in other lower-frequency VLA bands, we calculate the Q-band
luminosities,  $\nu L_{\nu}$ ($\equiv P_{45~{\rm GHz}}$) in units of erg~s$^{-1}$), which
covers the range $\sim 38.4< \log L_{45~{\rm GHz}}< 40.7$.  We
also use the core luminosity densities at 5 GHz,  $\nu L_{\nu}$ ($\equiv P_{5~{\rm GHz}}$ from \citet{kellermann89} which covers the range $\sim
37.6< \log P_{5~{\rm GHz}}< 40.1$.  Note that although we always
use the A-configuration for both the Q band and the C band
observations for the following radio luminosity plots (unless
explicitly expressed), the Q band resolution is 40-80 mas versus
$\sim$0.3 arcsec in the C band, due to the factor 9 drop in
frequency. Thus, if the radio emission is extended, the derived slope
between the two bands will be steeper than the intrinsic core slope. In this
scenario, since extended radio emission is almost always characterised by an 
optically-thin steep power-law, this possible bias will not
change the character of the 45-GHz emission.  This aperture bias does not
work in the opposite direction, and an observed flat spectrum is
inevitably flat (or even flatter if it is contaminated by a steep
component) and must therefore originate from an optically thick
compact source. An intrinsic compact source will remain unresolved in
both bands, and the measured slope will be unaffected by the different
resolutions.

Figure~\ref{multi_5_45} depicts the correlations of P$_{5~{\rm
    GHz}}$ (left panel) and P$_{45~{\rm GHz}}$ (right panel) with
the typical calorimeters of AGN strength: the X-ray (2-10 keV) luminosity
$L_{\rm X}$, the optical luminosity $L_{\rm o}$, the bolometric
luminosity $L_{\rm bol}$ and the [O~III] line luminosities $L_{\rm
  [O~III]}$.  Since several upper-limits are present in the
datasets, the statistical significance of each correlation is measured
by using a censored statistical analysis (ASURV\footnote{ASURV package
has been used within the PyRAF software \citep{pyraf2}.};
\citealt{lavalley92}) which takes into account the presence of upper
limits.  We used the \verb'schmittbin' task \citep{schmitt85} to
calculate the associated linear regression coefficients for two sets
of variables. Effectively, we carried out this procedure twice,
obtaining two linear regressions: first, we consider the former
quantity as the independent variable and the latter as the dependent
one and second switching the roles of the variables. The best fit is
represented by the bisector of these two regression lines. This
followed the suggestion of \citet{isobe90} that considers such a
method preferable for problems that require a symmetrical treatment of
the two variables. To estimate the quality of the linear regression,
we used the generalised Kendall's $\tau$ test \citep{kendall83} (task
\verb'bhkmethod') valid for sample smaller than 30 targets as in our
case and the associated probability that the correlation is
fortuitous. We have also measured the (maximum) rms of the
correlations, by assuming that censored data are
detections. Table~\ref{statistics} provides the statistical parameters, slope and intercepts of the linear regressions we evaluate in this
work.

\begin{table*}
\begin{center}
\caption{Statistical censored analysis of the tested correlations for RQQs and RLQs.}
\begin{tabular}{lllc|cccc|cc} 
\hline
\hline
    $X$  & $Y$ & Fig. & sample  & Stat & $\rho_{XY}$ & $P_{\rho_{XY}}$ &  rms & Slope  & Intercept  \\
    (1)   &  (2)   & (3)  & (4) & (5) & (6) & (7) & (8) & (9) & (10) \\
\hline
    $\log P_{5~{\rm GHz}}$ &   $\log L_{\rm X}$    & \ref{multi_5_45}      & RQQ & K & 0.857 & 0.0285 & 0.184 & 0.58$\pm$0.24  &  21.6$\pm$8.1  \\ 
    $\log P_{45~{\rm GHz}}$ &   $\log L_{\rm X}$    & \ref{multi_5_45}     & RQQ & K & 1.209 & 0.0011 & 0.117 & 0.60$\pm$0.20  &  20.5$\pm$7.6  \\ 
    $\log P_{5~{\rm GHz}}$ &   $\log L_{\rm o}$    & \ref{multi_5_45}      & RQQ & K & 0.933 & 0.0052 & 0.182 & 0.91$\pm$0.29  &  9.4$\pm$8.9  \\
    $\log P_{45~{\rm GHz}}$ &   $\log L_{\rm o}$    & \ref{multi_5_45}     & RQQ & K & 1.009 & 0.0046 & 0.172 & 0.90$\pm$0.29  &  9.0$\pm$6.1  \\
    $\log P_{5~{\rm GHz}}$ &   $\log L_{\rm bol}$    & \ref{multi_5_45}   & RQQ & K & 0.743 & 0.0260 & 0.233 & 0.97$\pm$0.32  &  7.8$\pm$9.1  \\
    $\log P_{45~{\rm GHz}}$ &   $\log L_{\rm bol}$    & \ref{multi_5_45}  & RQQ & K & 0.933 & 0.0081 & 0.168 & 0.99$\pm$0.34  &  6.7$\pm$9.5  \\
    $\log P_{5~{\rm GHz}}$ &   $\log L_{\rm [O~III]}$   & \ref{multi_5_45} & RQQ & K & 0.705 & 0.0436 & 0.250 & 0.816$\pm$0.25  &  11.4$\pm$13.6  \\
    $\log P_{45~{\rm GHz}}$ &   $\log L_{\rm [O~III]}$  & \ref{multi_5_45} & RQQ & K & 1.276 & 0.0003 & 0.238 & 0.811$\pm$0.19  &  11.0$\pm$12.5  \\
\hline
    $\log M_{\rm BH}$    & $\log P_{5~{\rm GHz}}$ &   \ref{multi_radio}     & RQQ & K & 0.5902 & 0.0746 & 0.469 & 1.17$\pm$0.35  &  29.7$\pm$19.1  \\
     $\log M_{\rm BH}$    &$\log P_{45~{\rm GHz}}$ &   \ref{multi_radio}     & RQQ & K &  1.0095 & 0.0023 & 0.297 & 1.17$\pm$0.28  &  30.3$\pm$33.7  \\
    $\log M_{\rm BH}$    & $\log P_{5~{\rm GHz}}$ &   \ref{multi_radio}      & RLQ & P & 0.576 & 0.066 &         \\  
     $\log M_{\rm BH}$    &$\log P_{45~{\rm GHz}}$ &   \ref{multi_radio}     & RLQ & K &  0.2167   & 0.5352 &    \\
\hline
     $\log H\beta \,\, FWHM$  & $\alpha_{5-45~{\rm GHz}}$    & \ref{multi_slope}     & RQQ & K &  0.8791 & 0.0258 &\\
      $\log H\beta \,\, FWHM$ & $\alpha_{5-45~{\rm GHz}}$    & \ref{multi_slope}     & RLQ & K & -0.850  & 0.0215 &  \\
      $\log L/L_{\rm Edd}$   &  $\alpha_{5-45~{\rm GHz}}$ & \ref{multi_slope}     & RQQ & K  & -0.593 & 0.132  &  \\
      $\log L/L_{\rm Edd}$  &   $\alpha_{5-45~{\rm GHz}}$ & \ref{multi_slope}     & RLQ & K &  0.0167 & 0.964  & \\
      $\log  M_{\rm BH}$    &  $\alpha_{5-45~{\rm GHz}}$ &\ref{multi_slope}     & RQQ & K &  0.725  &  0.0655 & \\
      $\log  M_{\rm BH}$    & $\alpha_{5-45~{\rm GHz}}$ & \ref{multi_slope}     & RLQ & K  & -0.567 & 0.125  & \\
\hline
  $\log L/L_{\rm Edd}$ &  $\log P_{45~{\rm GHz}} / L_{\rm o}$        & \ref{l45lxlo}  & RQQ & K & -0.991 & 0.0093 & 0.245 & -0.90$\pm$0.36  &  -5.5$\pm$4.1  \\
  $\log L/L_{\rm Edd}$&  $\log P_{45~{\rm GHz}} / L_{\rm o}$       & \ref{l45lxlo}     & RLQ & K & 0.0381 &  0.903  \\
  $\log M_{\rm BH}$   &  $\log P_{45~{\rm GHz}} / L_{\rm X}$    & \ref{l45lxlo}     & RQQ & K & 0.7619 & 0.0078 & 0.322 & 0.90$\pm$0.37  & -11.7$\pm$7.2 \\
  $\log M_{\rm BH}$   &  $\log P_{45~{\rm GHz}} / L_{\rm X}$    & \ref{l45lxlo}     & RLQ & K & -0.350 & 0.3387  \\    
\hline
\end{tabular} 
\label{statistics} 
\end{center} 
\begin{flushleft}
Column description: (1)-(2) the two variables of the considered
relation; (3) Figure; (4) sample (RQQ or RLQ) for the tested
correlation. (5)--(6)--(7) the statistical
analysis used for the given sub-sample to calculate the associated
linear regression coefficient $\rho_{\rm XY}$ and the probability that
there is no correlation $P_{\rho_{XY}}$: $K$ for the censored
generalised Kendall's $\tau$ correlation coefficient and $P$ for the
Pearson correlation coefficient for a fully detected data set; (8) the
rms scatter of the linear regression; (9)--(10) the slope and the
intercept of the best fits with their 1-$\sigma$ errors.
\end{flushleft}
\end{table*}

Figure \ref{multi_5_45} shows that P$_{45~{\rm GHz}}$ presents a
tighter correlations with $L_{\rm X}, L_{\rm opt}, L_{\rm bol}$ and
$L_{\rm [O III]}$ compared to those with P$_{5~{\rm GHz}}$. A clear
outlier from the X-ray--radio regressions is PG~1001+054, which is
known to be significantly X-ray absorbed \citep{schartel05}. Therefore
we excluded this target in the statistical analysis. From all
correlations presented in Fig.~\ref{multi_5_45}, the tightest fit is
P$_{45~{\rm GHz}}$-$L_{\rm X}$ (rms = 0.117) and the most
statistically significant one is P$_{45~{\rm GHz}}$-- $L_{\rm
  [O~III]}$ (P = 0.0003). The strength of these two relations is
remarkable, given the fact that the X-ray and optical nuclei are
generally more variable (variability amplitudes, typically, of a
factor of several, up to $\sim$50, e.g.
\citealt{markowitz04,paolillo04,guainazzi04,saez12,lanzuisi14}) than
those detected in the radio-band (variability amplitudes of a factor
of few, e.g. \citealt{barvainis05,mundell09}).

\subsection{Comparison with PG radio-loud quasars}
\label{comparison}

The PG quasar sample \citep{boroson92} also includes 16 RL AGN.  Below
we compare their 45~GHz properties with those of the PG RQQs to
search for significant similarities and differences between the two
quasar samples. This aims at investigating the nature of the radio
emission in RQQs in relation to the well-studied relativistic
collimated jet emission of RL AGN.

We obtain 45-GHz VLA observations in A and B configuration for 8 RLQs
from the literature (see Table \ref{lumRL}). For the remaining 8 sources,
the 45 GHz flux densities are roughly estimated by a power-law
extrapolation of their radio spectra (taken from the NASA/IPAC
Extragalactic Database\footnote{NED, http://ned.ipac.caltech.edu})
from lower frequencies\footnote{This method is supported by the fact
that all the RLQs are Fanaroff-Riley type-II radio galaxies, which
have powerful extended steep-spectrum jets which dominate the entire broadband radio
SED. As a counter-check, the 45-GHz VLA luminosities of the 8 RLQs
take from literature are consistent with the values extrapolated from
the data from NED within the scatter of the power-law relation.}. The typical
spectral coverage is from 74 MHz to 8.5 GHz and in some cases up to
the 31.4 GHz. The typical error bars on the 45-GHz flux densities for those
sources is estimated as a factor 2 of the measurements ($\sim$0.3 in
logarithmic scale).

\begin{figure}
\includegraphics[width=0.75\columnwidth,angle=90]{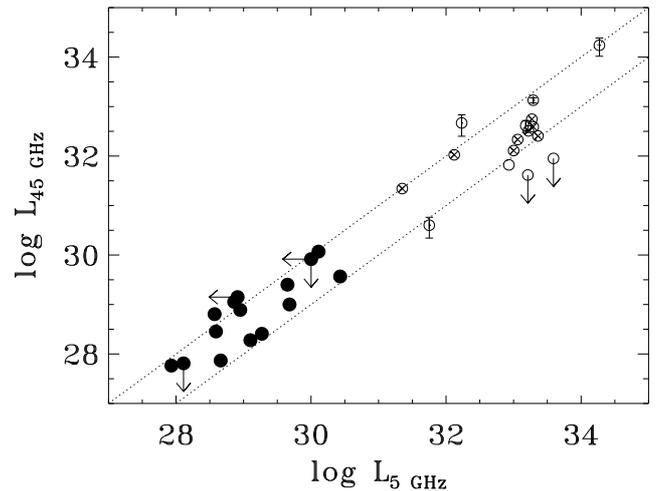}
\caption{The 5-GHz radio core luminosities vs the 45-GHz luminosity
  (erg s$^{-1}$ Hz$^{-1}$). The filled points are the 15 RQQs, the
  empty points are the RLQs with available VLA Q-band data and the
  crossed empty circles are the RLQs with the 45-GHz flux densities
  extrapolated from their radio spectra. The RLQs are typically $\sim
  1000$ more luminous than the RQQs at both 5~GHz and 45~GHz. The
  dotted lines are drawn to visualise a scatter of a factor of 10.
  The scatter of RQQs ($\sigma$=0.29) is smaller than that of RLQs
  ($\sigma$=0.41).}
\label{5_45}
\end{figure}

Figure~\ref{5_45} compares L$_{45~{\rm GHz}}$ versus L$_{5~{\rm GHz}}$
(erg s$^{-1}$ Hz$^{-1}$) of the PG RLQs and RQQs. The RLQs are
typically $\sim 1000$ times brighter than the RQQs at both 5~GHz and
at 45~GHz. The scatter of the relation between
 log L$_{45~{\rm GHz}}$ and log L$_{5~{\rm GHz}}$ of the PG RQQs
  ($\sigma$=0.29) is smaller than that for that RLQs ($\sigma$=0.41). This may result from enhanced core
  variability and orientation-dependent Doppler beaming or from the
  spectral extrapolation scatter, valid for RLQs.

\begin{table}
\begin{center}
  \caption[Properties of the RLQ sample.]{Radio properties of the PG RLQ sample.}
  \begin{tabular}{ll|cc}
    \hline
PG name & alternative   &   F$_{\rm core}$    &   $P_{45~{\rm GHz}}$ \\
\hline 
PG~0007+106 & MRK~1501  &   2.6$\pm$1.0$^{a,b,c}$ Jy  &    43.32  \\
PG~1226+023 & 3C~273    &  26.9$\pm$10.8$^{a}$  Jy   &     44.89  \\
PG~1302-102 & PKS~1302-102 & 0.59$\pm$0.07$^{a,d}$ Jy  &   43.78  \\
PG~1545+210 & 3C~323.1  &  $<$2.6$^{a}$ mJy           &     $<$42.27  \\
PG~1704+608 & 3C~351    &   $<$2.2$^{a}$ mJy          &    $<$42.61  \\
PG~2209+184 & II~Zw~171 &  37.6$\pm$17.2$^{a}$ mJy    &    41.25  \\
PG~1103-006 & PKS~1106-0052 &  67.65$\pm$14.4$^{e}$ mJy  &  43.27  \\
PG~1512+370 & 4C~+37.43 &  15.0$^{f}$ mJy                &  42.47  \\
\hline
PG~0003+158 & PKS~0003+15 & 56 mJy  &  43.25 \\
PG~1004+130 & 4C~+13.41   & 79 mJy  &  42.76 \\
PG~1048-090 & 3C~246      & 148 mJy &  43.40  \\
PG~1100+772 & 3C~249.1    & 110 mJy &  43.16  \\
PG~1309+355 &             & 25 mJy  &  42.00  \\
PG~1425+267 & B2~1425+26  & 25 mJy  &  42.68  \\
PG~2251+113 & 4C~+11.72   & 79 mJy  &  43.06  \\
PG~2308+098 & 4C~+09.72   & 34 mJy  &  42.99  \\
\hline
\hline                                          
\end{tabular}                                   
\label{lumRL}
\begin{flushleft}
  Column description: (1)-(2) PG source name and alternative name; (3)--(4)
  radio core flux densites  and luminosities  (erg s$^{-1}$) obtained from Q-band
    radio maps (upper list) and extrapolated by fitting the radio SED taken from NED
    (lower list) (See Sect 3.2 for details); references: $a$ NVAS (NRAO VLA Archive Survey), $b$
  \citet{gu09}, $c$ \citet{lanyi10}, $d$ \citet{perley17}, $e$
  \citet{sajina11}, $f$ \citet{bolton04}; (4) 45-GHz core luminosity
  (erg s$^{-1}$).  The typical error of the core flux densities of
  RLQs derived from their radio SED is approximately a factor $\sim$2 of the
  measurements.
\end{flushleft}
\end{center}
\end{table}

\begin{figure}
\centerline{\includegraphics[width=0.85\columnwidth,angle=90]{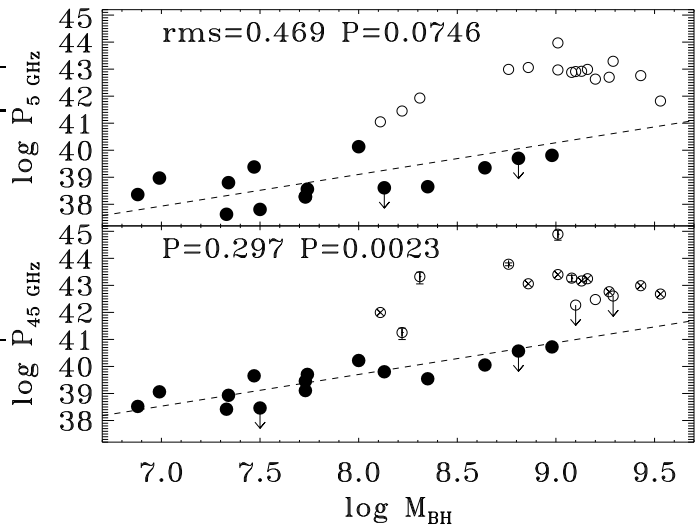}}
\caption{M$_{\rm BH}$ (in M$_{\odot}$) vs. P$_{5~{\rm GHz}}$
   (upper panel) and  vs. P$_{45~{\rm
      GHz}}$ (lower panel) in units of erg s$^{-1}$. The filled points are the 15 RQQs, the
  empty points are the RLQs with available VLA Q band data and the
  crossed empty circles are the RLQ with the 45-GHz flux densities
  extrapolated from their radio spectra.  The value of the generalised
  Kendall's $\tau$ test probability (P) for the RQQ relation and its
  rms is provided in each panel. Note the tight relation of P$_{45~{\rm GHz}}$ and M$_{\rm BH}$ valid for RQQs.}
\label{multi_radio}
\end{figure}

Since the BH mass has been found to be a crucial quantity for modeling
and interpreting observations in the framework of jet launching
mechanism of RL AGN \citep{blandford77,cattaneo09}, here we compare
the radio luminosities of the PG quasars with M$_{\rm BH}$.
Fig.~\ref{multi_radio} shows the dependence of $P_{45~{\rm GHz}}$ and
$P_{\rm 5GHz}$ on M$_{\rm BH}$ for the RL and RQ PG quasars.   The RQQs
show a strikingly tight linear correlations, (P=0.0023 for P$_{45~{\rm
    GHz}}$) with a smaller scatter at 45 GHz than at 5 GHz, in the
form P$_{45~{\rm GHz}} \propto M_{\rm BH}^{1.17}$, which is consistent
within the errors  with the slope of the M$_{\rm BH}$-radio
correlation established for nearby RQ AGN observed with eMERLIN
\citep{baldi21b}. For RLQs the radio-BH relations have significantly
larger scatter and are not statistically significant, with $P$ values
of only $\gtrsim$0.07. Apart from possible large systematic errors
due to the SED-extrapolated 45-GHz emission, the most plausible
physical reason of a larger scatter for the RLQ relation than that of
RQQs is the higher variability amplitude of RLQs due to Doppler
boosting.

\begin{figure*}
\includegraphics[width=1.4\columnwidth,angle=90]{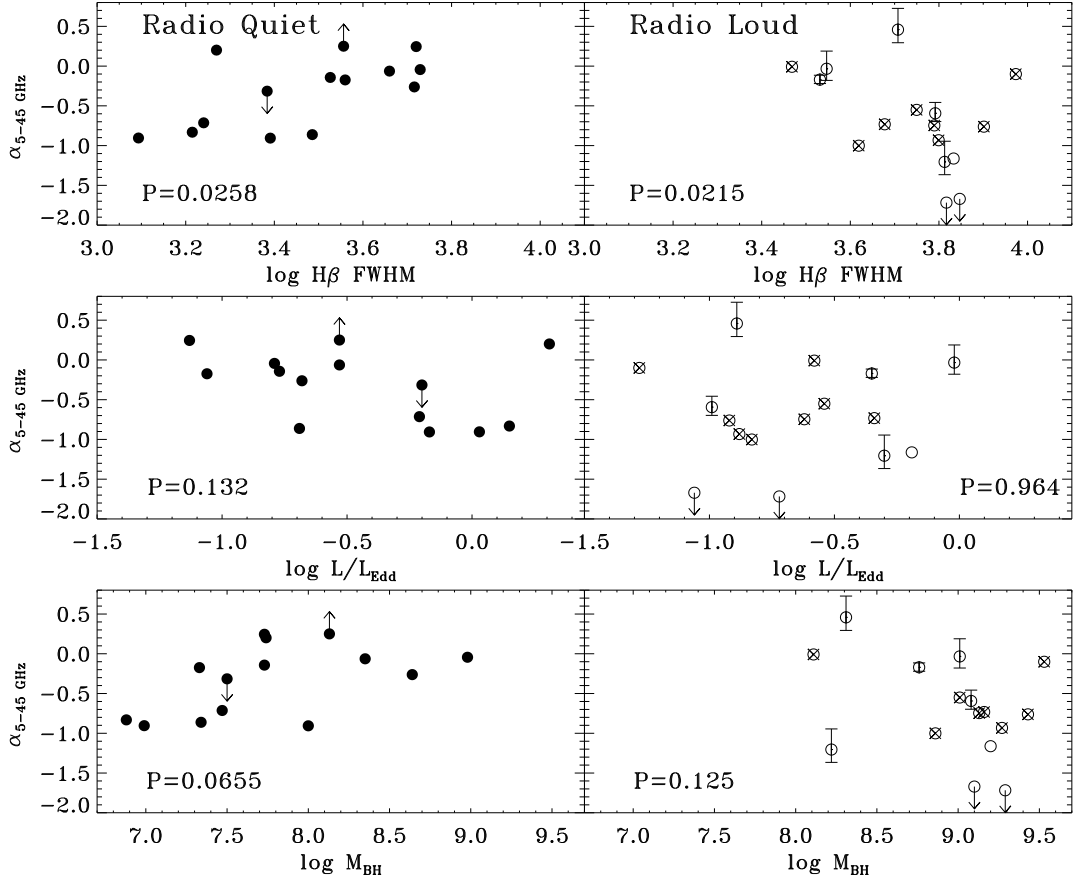}
\caption{The radio spectral index derived between 5 and 45 GHz as
  function of the H$\beta$ FWHM (km s$^{-1}$), L/L$_{\rm Edd}$ and
  M$_{\rm BH}$ (M$_{\odot}$).  The left panels show the 15 PG RQQs, whereas the
  right panels show the 16 PG RLQs. The filled points are the PG RQQs,
  while the empty points are the PG RLQs. Those with crosses are the
  PG RLQs whose 45-GHz radio luminosities have been estimated by
  fitting their radio SED. The value in the lower left corner of each
  panel expresses the generalised Kendall's $\tau$ test probability
  (P) of a fortuitous correlation and its rms. The strongest trends
  are with the H$\beta$ FWHM, but are opposite in RQQs and RQQs,
  suggesting different emission mechanisms.  }
\label{multi_slope}
\end{figure*}

To derive the spectral slopes between the C and Q bands (F$_{\rm \nu} \sim \nu^\alpha$), we used the $uv$-tapered Q-band core flux densities
which are extracted from the images with a resolution comparable with
the C band data, $\sim$0.3-0.5 arcsec.  The $\alpha_{5-45\rm{GHz}}$ values
range between -1 and 0.3 (Tab.~\ref{tab_alfa}, with typical errors of
30 per cent), a narrower interval than the spectral indices derived at
lower radio frequencies, 5-8.5 GHz \citep{laor19}. We compared the
5--45\,GHz indices with the parameters of the EV1 set of correlations
\citep{boroson92}, i.e., H$\beta$ FWHM, Eddington ratio L/L$_{\rm
  Edd}$, and M$_{\rm BH}$, in analogy to the analysis done by
\citet{laor19} (Fig.~\ref{multi_slope}, left panels). We fit the
censored data, and the derived relations between the radio slopes and
the three quantities for the RQQs appear to be statistically weak
(with probabilities of fortuitous correlations higher than 0.02, see
Tab~\ref{statistics}). General trends have been found: the radio slope
generally seems to flatten for larger H$\beta$ width, smaller
L/L$_{\rm Edd}$ and higher M$_{\rm BH}$, similarly to what was seen by
\citet{laor19}, but with lower statistical significance. For a
comparison with RLQs, we analogously fit their censored
$\alpha_{5-45\rm{GHz}}$ and EV1 quantities (Fig.~\ref{multi_slope}, right
panels). The statistics of the regressions are even lower than those
of RQQs, but we can conclude that RLQs generally display opposite
trends to the ones obtained with RQQs, which could slightly strengthen
the different radio physical properties of RQQs and RLQs. The
statistically strongest relation (P $\sim$ 0.02) is $\alpha_{5-45\rm{GHz}}$
-- H$\beta$ FWHM, where RQQs and RLQs clearly reveal opposite
regressions.

\begin{figure*}
\centering
\includegraphics[width=0.75\columnwidth,angle=90]{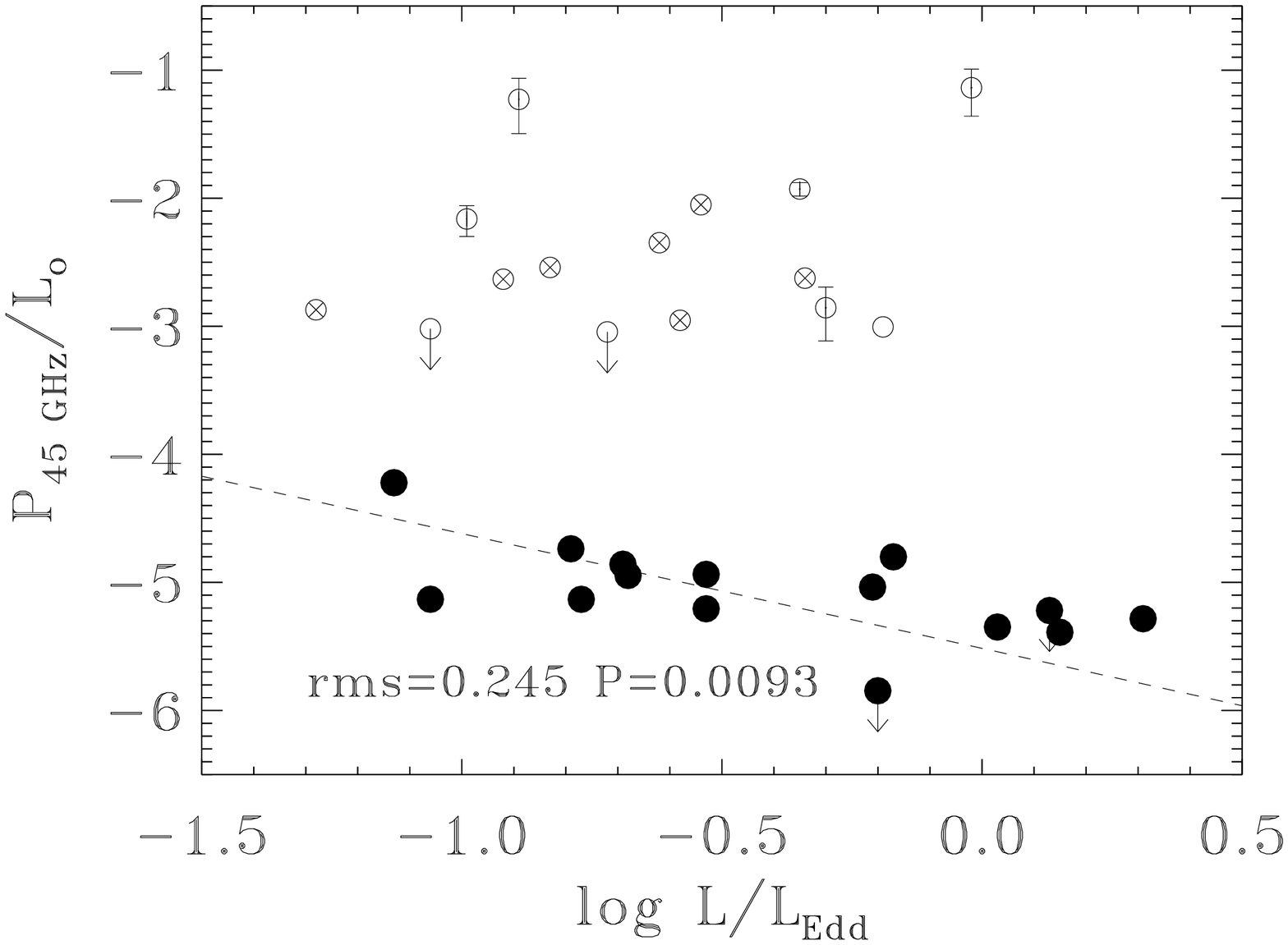}
\includegraphics[width=0.75\columnwidth,angle=90]{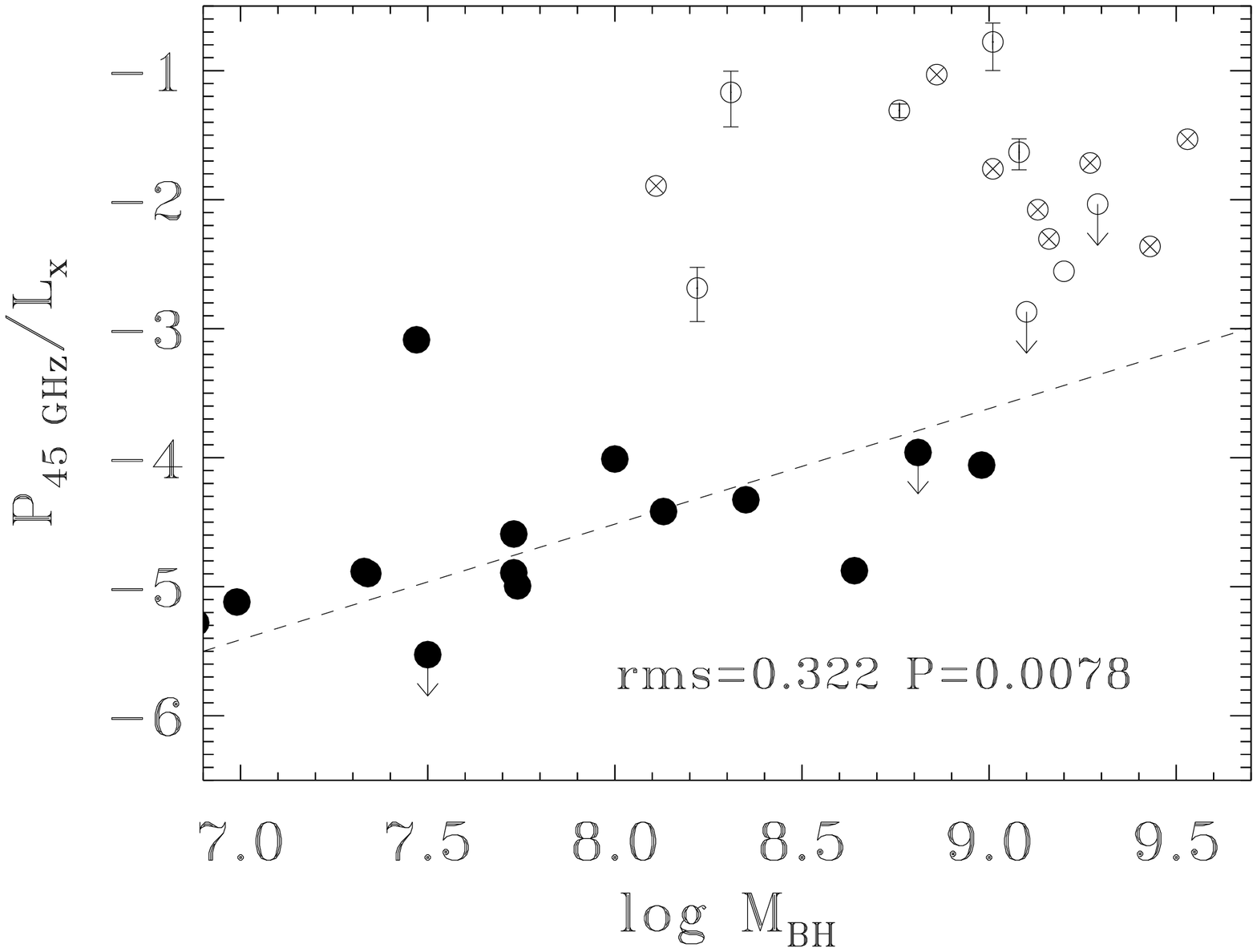}
\caption{Left panel: radio loudness estimated as ratio between 45-GHz
  and optical luminosities versus L/L$_{\rm Edd}$. Right panel: radio
  loudness estimated as ratio between 45-GHz and 2-10 keV X-ray
  luminosities versus M$_{\rm BH}$ (M$_{\odot}$). The filled points are the PG RQQs,
  while the empty points are the PG RLQs. Those with crosses are the PG
  RLQs whose 45-GHz radio luminosities have been estimated by fitting
  their radio SED. The value in the lower part of each panel expresses
  the rms and the generalised Kendall's $\tau$ test probability (P) of
  a fortuitous correlation for the RQQs. Note that the 45\,GHz-based
  radio-loudness decreases with increasing L/L$_{\rm Edd}$, while the X-ray based radio-loudness increases
  towards higher M$_{\rm BH}$, with a jump by a factor of $\sim$30 at
  $M_{\rm BH}>10^9$ M$_{\odot}$, where all the PG quasars become RL.}
\label{l45lxlo}
\end{figure*}

As the radio loudness has been related to the accretion properties and
BH mass in the past
(e.g. \citealt{nelson00,ho08,best05b,heckman14,chang21}), we also
investigate the 45\,GHz-based radio loudness of the PG quasars,
measured as P$_{45~{\rm GHz}}$/L$_{\rm o}$ and P$_{45~{\rm
    GHz}}$/L$_{\rm X}$, respectively as function of L/L$_{\rm Edd}$
and M$_{\rm BH}$ (Fig.~\ref{l45lxlo}). We find a tight P$_{45~{\rm
    GHz}}$/L$_{\rm o}$ -- L/L$_{\rm Edd}$ correlation (left panel, P=
0.0093, rms= 0.245), where the RQQs become quieter with increasing
L/L$_{\rm Edd}$.  At $L/L_{\rm Edd} \gtrsim$ 0.3, which is roughly the
value when the spectral indices steepen, $\lesssim$0
(Fig.~\ref{multi_slope}), we get that P$_{45~{\rm GHz}}$/L$_{\rm
  o}<10^{-5}$, the typical value expected for the corona disc
emission.  In addition we also find that P$_{45~{\rm GHz}}$/L$_{\rm
  X}$ is linearly correlated with M$_{\rm BH}$ (P=0.0078,
  rms=0.322): RQQs become louder with increasing M$_{\rm BH}$. Since
45-GHz luminosities have been found to correlate with the optical and
X-ray luminosities (despite not with a slope of unity), the dependence
of these luminosity ratios with L/L$_{\rm Edd}$ and BH mass highlights
a second-order effect in the radio emission production in RQQs. A
larger sample of RQQs would still be necessary to confirm this
result. Conversely, RLQs do not show statistically significant
correlations. However we can note that at M$_{\rm BH}>10^9$
M$_{\odot}$ all PG quasars become RL, and their P$_{\rm 45
  GHz}$/L$_{\rm X}$ jumps by a factor of $\sim 30$, compared to the
RQQs of our sample\footnote{The standard 5\,GHz-to-optical
radio-loudness parameter drops by a factor of $\sim$1000 from RLQs and
RQQs at M$_{\rm BH} \sim$ 10$^{9}$ M$_{\odot}$.}.

\section{Discussion}
\label{sec:discussion}

The overall 1.4 -- 45~GHz spectra of the 15 PG RQQs studied here
indicate that two main components shape the SEDs of most of the
sources: a steep-spectrum and a flat-spectrum component which
dominates, respectively, at low and high radio frequencies. The 45-GHz
emission is generally dominated by the latter, which in some objects
extends down to 5~GHz (at least in 9 sources), in particular in high
resolution observations.  A flat-spectrum core suggests an optically
thick source, which implies a physically compact source with a size
$\sim$0.001$-$0.02 pc (e.g. eq. 22 in \citealt{laor08}).

A compact source which characterises the 45-GHz emission, is also supported
by two further results from this study. In fact, most of our sources
are unresolved, i.e should be smaller than $\sim 50$~mas, which
corresponds to physical sizes smaller than $\sim 50-100$~pc. Only
  in two objects (PG~0844+349 and PG~1259+593) an unresolved core is
  not detected. These two sources also lack of further detections at
  high resolution and at lower radio frequencies, thus giving a
  coarse charaterisation of their radio spectra.

The other indirect evidence for the general compact nature of the
45~GHz emission comes from the tight correlation of $P_{45~{\rm GHz}}$
with the direct AGN luminosities $L_{\rm X}, L_{\rm o}, L_{\rm bol}$
and $L_{\rm [O~III]}$, in particular when compared with the
somewhat-weaker correlations of $P_{\rm 5 GHz}$ with these
quantities. This result is notable given the $\sim 20-30$ years time
span between the 45~GHz and the earlier optical and X-ray
observations, and is similar to previous studies that have found
striking radio--X-ray correlations at even smaller radio scales and at
lower luminosities (e.g \citealt{panessa13,panessa15}). As it occurs
at longer cm-band wavelengths, the 5~GHz luminosity may depend more on
the host galaxy properties, if it is produced e.g. by an AGN-driven
wind which shocks the host interstellar medium (ISM), or nuclear star
formation (e.g. \citealt{zakamska16,mancuso17,richards21}). Instead,
at higher radio frequencies, at 45~GHz in our case, the radio emission
may originate from an inner physical region at the accretion disc
scale, e.g. from coronal disc emission, or a weak jet
(e.g. \citealt{chiaraluce20,smith20b}, see Sect.~\ref{origin} for the
discussion on the radio origin), and is set by the AGN
properties.

The genuine compact nature of the 45~GHz emission of the PG RQQs which
clearly show spectral flattening at high frequencies, could be tested
using VLBI observations. A few available VLBI studies of RQ AGN
(e.g. \citealt{blundell98,ulvestad05_rqq,doi13}) reveal a mas-scale
($\sim pc$) flat-spectrum compact source, which extrapolates well to
the observed higher radio-frequency emission \citep{laor19}. Future
VLBI observations of the PG RQQs, part of our survey, will confirm their
compactness.

Although not statistically very robust, the suggested relations
between $\alpha_{\rm 5-45}$ and the H$\beta$ FWHM, L/L$_{\rm Edd}$ and
M$_{\rm BH}$ are consistent with the results found by \citet{laor19}
for the PG quasars by using radio slopes at 5--8.5~GHz.  The most
significant relation is the the steepening of the radio slopes of RQQs
with L/L$_{\rm Edd}>0.3$ and low H$\beta$ widths. This Eddington ratio
limit can be interpreted at the condition of a radiatively efficient
accretion disc, required to launch a radiation pressure driven wind,
which then shocks the host ISM and produces a steep-spectrum extended
synchrotron source \citep{king13,zakamska14,nims15}. Conversely, the
dependence on H$\beta$ FWHM is not simply interpretable as a
orientation effect, as a flatter spectra would be expected in a polar
view. The weaker correlations found here than the ones found with
$\alpha_{\rm 5-8.5}$ \citep{laor19} may result from the fact that
$\alpha_{\rm 5-45}$ is more an estimate of the relative contributions
of the steep and flat sources. In contrast, the radio slope at
5-8.5~GHz, is a more physical estimate of the nuclear properties and
likely a more accurate measure of the outflow characteristics.

A remarkable result of this study is the presence of
  unprecedentedly tight correlation between the 45-GHz radio
  luminosity and the BH mass for our sample, with a dependence similar
  to what has been found for local low-luminosity RQ AGN, but with a
  larger scatter \citep{baldi21b}. The connection between the
  mas-scale radio properties (luminosity, spectral slope, radio
  loudness) and the BH mass for our sample suggests that the latter
  should play an important role at setting the mechanism of radio
  production in RQQs (in jet and wind scenarios). In addition, the
  radio loudness parameter has been found to increase with BH mass and
  decrease with the Eddington ratio, consistent with previous studies
  (e.g. \citealt{ho02,merloni03,nagar05,sikora07,gurkan19,chiaraluce20}). However,
  in support of a distinct origin of the radio emission between RLQs
  and RQQs is the different (even opposite) trends with radio slopes
  and BH mass. Therefore, even in the same quasar regime (at high
  Eddington ratios), the radio outflow of RQQs must differ in terms of
  its driving mechanism compared to the strong jet observed in RLQs.

Based on the statistical significance of the tested correlations, the
45-GHz emission in PG RQQs is mainly driven by the X-ray radiation and
the BH mass. The X-ray emitting corona coincide with the jet base for
RLQs \citep{markoff05} and with the physical region where 45-GHz
emission is produced, more internal than the 5-GHz core, for RQQs
\citep{laor08}.  The similarity of the 45-GHz emitting region of RLQs
and RQQs would possibly suggest that the corona is playing an
important role in the radio emission production in quasars and the
properties of such a corona could account for the transition from a
relativistic jet in case of RLQs to a sub-relativistic outflow in case
of RQQs at high radio frequencies.

\subsection{Origin of the 45-GHz emission in RQQs}
\label{origin}

Here we explore the various possible scenarios of radio-emitting physical mechanisms
(see \citealt{panessa19}), which may account for the results we find
in this work.

\begin{itemize}
\item {\it Star formation}. A compact nuclear starburst could be
  scarcely consistent with the flat-spectrum unresolved cores we
  find. High star formation (SF) rate is generally expected at high
  accretion-rate AGN, as expected in a quasar regime
  \citep{sani10}. At 45 GHz, in this scenario the radio emission could
  primarily powered by free-free emission from discrete H~II regions
  on sub-kpc scale, making it an excellent tracer of massive SF
  \citep{murphy18}. The expected SF rate estimated in this radio band
  is $\sim$0.6--60 $M_{\odot}$ yr$^{-1}$ (equation 15 from
  \citealt{murphy11}, assuming electron temperature of 10$^{4}$ K and a
  flat-spectrum non-thermal component), which is not far from what has
  been found in luminous quasar hosts \citep{jarvis20,shangguan20} but
  is generally considered insufficient to explain the observed radio
  emission from quasars by an order of magnitude
  \citep{zakamska14,zakamska16}. Nevertheless, the morphological compactness of
  the radio emission and the multi-band correlations of the radio
  luminosities with AGN parameters (M$_{\rm BH}$, $L_{\rm bol}$,
  L/L$_{\rm Edd}$) argue against SF as the origin of the
  high-frequency radio emission in PG RQQs.

\item {\it Disc wind}. An AGN-driven wind interacting with the ISM and
  the consequent shock acceleration can cause synchrotron emission
  (see models from \citealt{jiang10,nims15}). The optically-thin radio
  emission is expected from an outflowing plasma from a standard
  accretion disc \citep{shakura73} with high radiation pressure,
  generally related to a high L/L$_{\rm Edd}$.  Such scenario would
  explain the relation between steep spectral slopes and the high
  L/L$_{\rm Edd}$ and large bolometric luminosities. A faster wind
  being accelerated by higher continuum-scattering (i.e. higher
  Eddington ratio, \citealt{gofford15}) will produce stronger shocks
  which will accelerate more electrons, resulting in higher radio
  luminosities. The most luminous radio sources would approximately
  lie on the relation between the radio luminosity and the [O~III]
  line width established by \citet{zakamska14}\footnote{To be
  consistent with the line width -- $L_{\rm 1.4 GHz}$ relation
  introduced by \citet{zakamska14} for obscured quasars, we derive the
  velocity width containing 90 per cent of [O~III] line, by assuming
  the conservative scenario where line is blueshifted and [O~III] and
  H$\beta$ widths are similar (H$\beta$ are slightly [8 per cent]
  systematically narrower than [O~III] on average,
  \citealt{zakamska14}). In addition, we derive the 1.4-GHz
  luminosities from our measurements, assuming a flat-spectrum
  synchrotron component with a slope of 0.5. Therefore, this method
  with several assumptions provides an approximate estimate of the
  radio emission from a disc wind model for our sample.}, interpreted
  as synchrotron radio emission form a quasar-driven winds propagating
  into the ISM of the host galaxy.  A steep radio spectra would be
  consistent with synchrotron emission from a diffuse disc wind on
  the scale of the galaxy. Moreover, the radio production based on
  this mechanism is not predicted to depend directly on M$_{\rm BH}$,
  in contrast with our results. In conclusion, we cannot fully rule
  out the possibility that the 45-GHz radio luminosities of the
  high-Eddington luminous PG RQQs have a significant contribution from
  a disc wind.

\item {\it Jet}. An uncollimated sub-relativistic jet, whose power is
  scaled down with respect to the jets in RLQs, is another possible
  scenario. As already discussed by \citet{laor19}, an orientation
  dependence of the radio properties similar to that interpreted for
  the RLQ unification scheme would lead to an opposite relation
  between the observed spectral slope and the H$\beta$ width: in the
  case of a type-1 AGN by definition, low inclination would tend to
  show smaller line widths and flatter radio spectra (an
  isotropically-emitting outflow could still be possible). The
  observed M$_{\rm BH}$ correlation with the radio luminosity and
  radio-loudness is consistent with the idea of a jet lunching
  mechanism by extracting energy from a spinning supermassive BH,
  specifically the Blandford-Znajek process \citep{blandford77}. A
  large BH masses and low Eddington ratios a core-brightened jet
  component is possibly emerging in the radio band, in agreement with
  what has been found for local Seyferts \citet{baldi21b}.  However,
  we do not find any continuity of the radio properties between the
  RQQs and the RLQs of the sample\footnote{It is still not clear
  whether a bimodal or continuous radio-loudness distribution of RQ
  and RL AGN would be distinct in the radio properties of the two
  classes (in terms of e.g. kinetic luminosity, size, speeds) (e.g.,
  \citealt{kellermann89,cirasuolo03,ho08,macfarlane21})}, excluding
  any possible result bias due to the incompleteness of the sample or
  because they come from the same original PG sample. Furthermore, the
  RLQs reveal a hint of reverse trends to the radio-based
  correlations shown by RQQs, suggesting different driving mechanisms
  of radio production.

\item {\it Coronal wind}. A magnetically-active corona linked to an
  accretion disc can produce synchrotron radio emission from the
  accelerated relativistic electrons produced by magnetic reconnection
  \citep{laor08,raginski16}. A dependence of the radio production on
  the X-ray power, produced by the cooling electrons, and the
  Eddington ratio is expected. In fact the high-frequency radio
  emission tightly correlates with high-energy nuclear component,
  suggesting a causal common emitting region, i.e. corona.  The
  observed radio to X-ray luminosity ratio agrees with the 10$^{-5}$
  ratio found for the analogous active coronae in cool active stars
  \citep{guedel93}. Highly active coronae, associated with high
  L/L$_{\rm Edd}$, are able to produce extended radio emission with an
  optically-thin bubble emerging from the outflowing corona similar to
  a Coronal Mass Ejection.  At higher X-ray luminosities, the
  height of the corona would increase \citep{alston20} with a higher
  probability of outflowing. At low L/L$_{\rm Edd}$, the radio
  emission is only related to the flat-spectrum self-absorbed emission
  emanated by an unresolved corona. When the Compton scattering is the
  dominant cooling mechanism, the luminosities from the corona should
  follow $\propto$ M$_{\rm BH}$ \citep{taam08,liu09}, consistent with
  the present results at 45\,GHz.

\end{itemize}

Based on our result, the most probable scenarios to account for the
radio properties of PG RQQs is that their radio cores originate from
non-thermal relativistic electrons likely accelerated in an active
corona, where intense magnetic reconnection events occur. However, a
possible increment of the emission contribution from a weak jet, as
the BH mass increases, cannot be ruled out, similarly to the jet power
dependence on M$_{\rm BH}$ observed in RL AGN \citep{liu06}, although the
role of such a quantity in characterising the RQ AGN properties is still controversial
(e.g.,
\citealt{boroson02,metcalf06,gurkan19,macfarlane21,baldi21b}). In our
case, this interpretation will accommodate the flat-spectrum cores we
observe at low L/L$_{\rm Edd}$ and large BH masses. In addition, the
most luminous quasars at high Eddington ratios could have an important
contribution in the radio band from disc winds. In fact, when the AGN
disc is able to drive strong winds due to high radiation pressure,
optically-thin intense steep-spectrum radio emission can be emanated
either by the accretion disc itself or from the outflowing corona. The
absence of such outflowing component at low L/L$_{\rm Edd}$ is due to
the decreasing of the radiation pressure and thus its relative radio
contribution, which make the flat-spectrum core emerges over the
steep-spectrum component. Nevertheless, it appear clear that a nuclear
star-forming core does not appear to reconcile with our results and
with a general interpretation of the radio emission in type-1 QSOs
\citep{zakamska16}.

\subsection{Comparison with X-ray Binary Systems}
\label{XRB}

Since jets are observed across all types of active BHs, it is
interesting to ask whether there are any similarities between the
radio properties of the RQQs and those of X-ray binary systems (XRBs)
and hence whether RQQs correspond to any particular X-ray 'state' of
XRBs. \citet{fender04} provide a full description of the change of
radio luminosity with 'state'. In the low/hard X-ray state where, very
broadly, the accretion rate is low or only moderate and the accretion
is probably radiatively inefficient, there is steady and relatively
powerful radio emission from a jet, with radio and X-ray luminosities
being well correlated (e.g. \citealt{corbel13}). Although not the only
possible interpretation, the correlations shown in
Fig.~\ref{multi_5_45} are quite consistent with such a scenario. At
very high accretion rates, i.e. the soft X-ray state where the
accretion is radiatively efficient, there is no detectable radio
emission. For example, the persistent soft-state XRB 4U1957+11
\citep{maccarone20} is undetected in radio maps with noise level of
1.07 $\mu$Jy/beam, whilst at a 2-20 keV X-ray flux of 10$^{-9}$ erg
cm$^{-2}$ s$^{-1}$. The radio/X-ray ratio for the present PG RQQs is
10$^{-5}$ far greater than value for XRBs.

Part of the increased radio emission in RQQs with respect to XRBs may
be due to the BH mass dependency in core-dominated
jets. \citet{heinz03} argue that AGN should be 3 -- 4 orders of
magnitude more radio loud than XRBs. However, with radio flux
densities 3 or 4 orders of magnitude above the upper limit for
4U1957+11, and X-ray fluxes about 2 orders of magnitude less than that
of 4U1957+11, the PG RQQs are still at least 1-2 orders of magnitude
more radio loud than expected if they were soft state XRBs, even
taking account of the mass dependency. Consequently, RQQs are probably
not simply soft-state systems (see \citealt{davis20}).

The increase of radio luminosity with BH mass seen in our sample is
consistent with at least a contribution from a compact jet in the
radio emission in a hard-state scenario. Conversely, a decrease in
radio luminosity with increasing accretion rate is not consistent with
a jet scenario \citep{heinz03}, at least if the RQQs remain in the
hard state.  However, the variation of radio luminosity with accretion
rate and BH mass in RQQs, differently from XRB states, likely
indicates a complexity of different accretion-ejection modes
(e.g. \citealt{fernandez21,baldi21b}) and that at least two sources of
radio emission co-exist for RQQs.

\section{Conclusions}
\label{concl}

This initial study (15 targets) of a comprehensive radio survey of
the PG RQQs (71 objects), named PG-RQS, reveals the following main results.

\begin{enumerate}

\item The 45-GHz emission is detected in 13 out of 15 PG RQQs, is generally
  unresolved, implying a source smaller than typically 50-100~pc.

\item The high frequency emission in PG RQQs is generally
  characterised by a flat spectrum component, that indicates a compact
  optically thick source.

\item The 45-GHz luminosity is tightly correlated with the AGN nuclear
  strength (X-ray, optical, [O~III], bolometric luminosities), suggesting an origin
  at the accretion disc scale. The 45-GHz luminosity is also
  correlated with BH mass.

\item The radio spectral properties of RQQs marginally correlate with
  AGN parameters: a flat-spectrum core emerges at low Eddington
  ratios, large BH masses and small H$\beta$ widths. 

\item The 45~GHz-based radio loudness decreases with increasing Eddington
  ratio and decreasing BH mass.

\end{enumerate}

The most plausible scenario to account for the general properties of
the compact 45-GHz cores detected in our sample is synchrotron
emission from a magnetically active X-ray coronae. However, at high
Eddington ratios and large BH masses a radio-emission contribution
from a disc wind and weak jet, respectively, cannot be ruled out.
These preliminary intriguing results need to be verified by extending
this study to the complete sample of the 71 PG RQQs, which will help
us to improve the statistical significance of the weak regressions
found in this work. Further studies, within the PG-RQS survey, at
higher resolution using the VLBI technique, together with an expanded
spectral coverage from low frequencies using {\it GMRT} and {\it
  LOFAR} observatories, to the mm and sub-mm regime using {\it ALMA},
will likely allow to determine the range of radio emission mechanism
in RQ AGN, and shed new light on a host of physical processes from
star formation on kpc scale down to the coronal structure on milli-pc
scale. The future step will be the advent of {\it SKA}, which will
survey vast numbers of local RQ active galaxies with $\mu$Jy-sensitivity at low
and intermediate radio frequencies \citep{orienti15}, providing the
cornerstone of our understanding of radio emission across a wide range
of galaxy types and accretion properties.


\section*{Acknowledgments}

We thank the anonymous referee for his/her helpful comments to improve
the manuscript.  A.H. acknowledges support by the I-Core Program of
the Planning and Budgeting Committee and the Israel Science
Foundation, and support by ISF grant 647/18.  This research was
supported by Grant No. 2018154 from the United States-Israel
Binational Science Foundation (BSF). A.L. acknowledges support by the
Israel Science Foundation (grant no. 1008/18). E.B. is funded by a
Center of Excellence of THE ISRAEL SCIENCE FOUNDATION (grant
No. 2752/19). I.M. acknowledges support from STFC under
ST/R000638/1. The National Radio Astronomy Observatory is a
  facility of the National Science Foundation operated under
  cooperative agreement by Associated Universities, Inc.

\section*{Data availability}

The radio data used in this work have been obtained by the VLA. The
uncalibrated dataset is public and available from the VLA data
archive (project 16B-126). Calibrated image products are available upon reasonable request to the corresponding author.

\bibliography{my}

\label{lastpage}
\end{document}